# Predicting Tourism Demand in Indonesia Using Google Trends Data

Atika Nashirah Hasyyati; Rina Indriani; Titi Kanti Lestari

BPS – Statistics Indonesia

## Abstract

Tourism data is one of the strategic data in Indonesia. In addition, tourism is one of the ten priority programs of national development planning in Indonesia. BPS-Statistics Indonesia has collected data related to tourism demand in Indonesia, but these data have different time period. Several data can be provided monthly, while the other data can be provided annually. However, accurate and real time tourism data are needed for effective policy making. In this era, all of information about tourism destination or accommodation can be gotten easily through internet, especially information from Google search engine, such as information about tourism places, flights, hotels, and ticket for tourism attractions. Since 2004, Google has provided the information of user behavior through Google Trends tool. This paper aims to analyze and compare the patterns of tourism demand in Indonesia from Google Trends data with tourism statistics from BPS-Statistics Indonesia. In order to understand tourism demand in Indonesia, we used Google Trends data on a set of queries related to tourism. This paper shows that the search intensity of related queries provides the pattern of predicted tourism demand in Indonesia. We evaluated the prediction result by comparing several time series models. Furthermore, we compared and correlated the Google Trends data with official data. The result shows that Google Trends data and tourism statistics have similar pattern when there were disasters. The result also shows that Google Trends data has correlation with official data and produced accurate prediction of tourism demand in Indonesia. Therefore, Google Trends data can be used to predict and understand the pattern of tourism demand in Indonesia.

# I. Introduction

The crucial issue that is related to tourism data in Indonesia is the gaps of providing tourism data. It is because of the differences in time period that BPS-Statistics Indonesia can provide these kinds of data. International visitor arrivals can be provided monthly, but domestic tourist data can only be provided annually. Data of international visitor arrival in Indonesia is obtained from Immigration Office and Mobile Positioning Data (MPD), while domestic tourist data is obtained from survey conducted by BPS-Statistics Indonesia. However, tourism data is very crucial to be provided in real time in order to fulfill the indicators for international and national development planning.

In national development, tourism has become the leading sector. Indonesian Government has committed to encourage the development of tourism, so it can be competed with other countries (MoT, 2017). Tourism also has an important role in the achievement of the 2030 Agenda of Sustainable Development Goals (SDGs). It can accelerate the achievement of all goals in the agenda because it has multiplier effect on other sectors. The contribution of tourism can be directly or indirectly to all of the goals, specifically it has been included in Goals 8, 12, and 14 on inclusive and sustainable economic growth, sustainable consumption and production (SCP) and the sustainable use of oceans and marine resources (UNWTO, 2015).

In order to achieve SDGs' Goals, Indonesian Government have ten national priority programs listed in Presidential Decree on Government Work Plans in 2018. One of them is focusing on tourism. This program will develop three priority tourism destinations, such as Toba Lake, Borobudur, Mandalika and the surrounding. These destinations will be introduced as "New Bali". Moreover, this program also develops tourism industries, tourism investment, jobs, and services export.

In this era, people can easily get a lot of information through internet, especially from Google Search. They can search information about travel planning, such as tourism destination, hotels/accommodations, flights, tourism attraction, and other information related to tourism. Individual who has a planning for vacation may search for places of interest, airline tickets, or the price of hotel rooms (Goel, 2010). Google Trends provides information about visitor behavior at the local or destinations level where it is not captured by existing surveys. The information from Google Trends is extremely up to date as it provides weekly figures for a period up to and including the current (incomplete) week. The availability of data from 2004 onwards allows a time-series to be built up relating to particular search terms (Smith, 2011).

Government agencies periodically release indicators of the level of economic activity in various sectors. However, these releases are typically only available with a reporting lag of several weeks and are often revised a few months later. It would clearly be helpful to have more timely forecasts of these economic indicators (Choi and Varian, 2011). Based on the 2018 Indonesian Government Work Plans, statistical data policy is directed to create the availability of data and more qualified statistical information that fulfil the criteria in accuracy, relevant, actual, timeliness, accessibility, and coherent to support the planning and evidence-based policy. Therefore, policy makers need statistical data with those criteria. Better quality of data are also needed to monitor and evaluate the sustainable tourism in the 2030 Agenda. There are several issues that need to be solved by National Statistics Office (NSO) such as shorten data lag, more detailed data, more frequent data availability, reduce outlier of missing data, and data standardization for official statistics (Bank Mandiri, 2017).

Google Trends offers more accurate forecasting, particularly for tourism. Predictions based on Google searches are advantageous for policy makers and business operating in the tourism sector (Zeynalov, 2017). Because of the timeliness of Google Trends, there has been a range of studies that examine how the data can be used to monitor economic trends. This avoids the time lag that is a feature of official statistics releases (Smith, 2011).

This paper will explore about visitor behavior from internet search to investigate whether Google Trends can show the pattern of tourism demand in Indonesia or not. In other words, this paper studies the appropriateness of Google Trends to be a complement for tourism statistics in order to close the gaps in economic statistics particularly tourism statistics for sustainable development. This paper also studies the prediction of tourism demand in Indonesia. It can be used for decision making about tourism development in destinations, industries, or infrastructure in Indonesia.

## II. Predicting Tourism Demand in Indonesia

### A. Google Trends Data for Tourism Statistics

#### 1. Literature Review

In today's, information gathering often consists of searching online sources. Recently, the search engine Google has begun to provide access to aggregated information on the volume of queries for different search terms and how these volume change over time, via the publicly available service Google Trends (Preis, 2013).

Ettredge (2005) investigates the potential of using data about web searches to predict an important macroeconomic statistic, specifically the number of unemployed workers in the U.S. In this article, rates of unemployment-related searches by Internet users are associated with unemployment levels disclosed by the U.S. government in subsequent monthly reports. A positive, significant association is found between the job-search variables and the official unemployment data. Smith (2011) studied the potential for Google Trends to provide information about visitor behavior at the local or destination level where it is not captured by existing surveys. It confirmed that information from Google Trends can be used to supplement official tourism statistics but that care is needed if it is to be used as proxy for these.

Choi and Varian (2011) have shown that data from Google Trends can be linked to current values of various economic indicators, including automobile sales, unemployment claims, travel destination planning, and consumer confidence. In 2009, Choi and Varian already reported some simple forecasting methods (AR models and fixed-effects models) using Google Trends data. Xin Yang et.al. (2014) forecasted Chinese tourist volume with search engine data (compared Google and Baidu), they verified the co-integration relationship between search engine query data and visitor volumes to Hainan Province. The study also demonstrated the value of search engine data, proposed a method for selecting predictive queries, and showed the locality of the data for forecasting tourism demand (Yang et.al., 2014).

Davidowitz and Varian (2015) wrote A Hands-on Guide to Google Data which covered three data sources (Google Trends, Google Correlate, and Google Consumer Survey). The author also explained that Google Trends provides an index of search activity on specific terms and categories of terms across time and geography. Google Correlate finds queries that are correlated with other queries or with user-supplied data across time or US state, meanwhile, Google Consumer

Surveys offers a simple, convenient way to conduct quick and inexpensive surveys of internet users (Davidowitz and Varian, 2015).

Chamberlin (2010) investigated the use of Google Trends data for various search categories, looking at its correlation with official data on retail sales, property transactions, car registrations and foreign trips, he concluded that Google Trends data may also be used informally to pick up on changing patterns and rising trends in search queries and the implications they have for types of economic activity and spending. Office for National Statistics (ONS) in 2011 explored the potential of using data on the internet search behavior of visitors to inform strategic decision making relating to tourism destinations. They concluded that Google Trends provides an effective measure of levels of interest in one or more topics and of changes over time in this interest. In India case, the paper of Mitra, Sanyal, and Choudhury (2017) attempts to bridge the data gap by nowcasting growth of real estate using Google Search Data. Their paper observes that the search intensity of relevant keywords unveils the demand condition of real estate sector on real time basis and using search intensity data provides better precision than other benchmark approaches for one quarter ahead forecast (Mitra, Sanyal, and Choudhury, 2017). Meanwhile, Anirbal Sanyal and Ira Irfan (2017) tracked labour market condition using Google Search data in India by developing composite indices and found that the indices corroborate Okun's law.

However, there are several weaknesses and considerations of using Google Trends. The conclusions about interest within sub-regions based on the data have to be explained carefully because Google uses IP address information. When using this tool in a tourism context include the fact that users of Google may not be representative of all visitors to a destination, for example, people may be more likely to use a travel agent or may not have internet access or less. In addition, people use may use search tools to a lesser extent, for example, regular visitor to a destination or visiting friends and families (Smith, 2011).

## 2. Data

Google Trends data related to tourism demand was used. Google Trends data was compared with official data released by BPS-Statistics Indonesia. Google Trends data to predict the pattern of international tourism demand was extracted from 1 January 2004 to 31 December 2017. Meanwhile, to predict the pattern of domestic tourism demand, Google Trends data from 1 January 2004 to 31 December 2016 was extracted. In this case, official data that has been used are number of monthly international visitor arrival data (January 2004 – December 2017) and annually domestic tourist data (2004 – 2016).

## 3. Methodology

### a) Google Trends Data

Google trends is a tool provided by Google in order to get information about internet searches with several filtering options, such as period, geographical area, type of information search (arts and entertainment, autos and vehicle, travel, etc.), comparison of queries search hits, related queries, feature for downloading Google Trends data.

Google "Insights for Search" analyses a portion of Google web searches to compute the number of searches that have been carried out for specific terms, relative to the total number of searches for the same term on Google over time (ONS, 2011). The data in "Insights for Search" are displayed on a scale of 0 to 100 after normalization, and each point on the graph has been divided by the highest point (ONS, 2011). Normalization means that Google has divided sets of data by a common variable to cancel out the variable's effect on the data, so this ensures that the underlying characteristics of the data sets can be compared (ONS, 2011). There are three methods of using Google Trends to make comparisons and each of these has filters that improve the analytical capacity of the tool (ONS, 2011):

1. Focus on one or more search term and filter by type of search, location of person making the search, date and category.
2. Alternatively, it is possible to select one or more locations of people making searches, compare the interest in a specific search term in these places and filter by type of search, date and category.
3. Finally, users can choose one or more-time range (e.g. individual years or months), compare the interest in a search term in these periods and filter by type of search, location and category.

### b) Bayesian Structural Time Series (BSTS) Model

Bayesian estimation and inference has a number of advantages in statistical modelling and data analysis, for instance, the Bayes method provides confidence intervals on parameters and probability values on hypotheses that are more in line with commonsense interpretations (Congdon, 2006). Andy Pole, Mike West, and Jeff Harrison (1994) stated that Bayesian statistical analysis for a selected model formulation begins by first quantifying the investigator's existing state of knowledge, beliefs, and assumptions. These prior inputs are then combined with the information from observed data quantified probabilistically through the likelihood function – the joint probability of the data under the stated model assumptions. The resulting synthesis of prior and likelihood information is the posterior distribution or information. The posterior is proportional to the prior and the likelihood,

$$posterior \propto prior \times likelihood$$

the prior to posterior process is referred to as Bayesian learning.

Peter Congdon (2006) stated that the learning process involved in Bayesian inference is one of modifying one's initial probability statements about the parameters before observing the data to updated or posterior knowledge that combines both prior knowledge and the data at hand. Bayesian models are typically concerned with inferences on a parameter set $\theta(\theta_1, \ldots, \theta_d)$ of dimension d. Prior knowledge about the parameters is summarized by the density $p(\theta)$, the likelihood is $p(y|\theta)$, and the updated knowledge is contained in the posterior density $p(\theta|y)$. From the Bayes theorem

$$p(\theta|y) = p(y|\theta)p(\theta)/p(y),$$

where the denominator on the right side is the marginal likelihood $p(y)$. The latter is an integral over all values of $\theta$ of the product $p(y|\theta)p(\theta)$ and can be regarded as a normalizing constant to ensure that $p(\theta|y)$ is a proper density. Then, Bayes theorem can be expressed as

$$p(\theta|y) \propto p(y|\theta)p(\theta)$$

Brodersen et.al. (2015) stated that structural time series models are useful in practice because of its flexibility and modularity. Structural time series models are state space models for time series data and can be defined as (Brodersen, et.al., 2015)

$y_t = Z_t^T \alpha_t + \varepsilon_t$     as observation equation,

$\alpha_{t+1} = T_t \alpha_t + R_t \eta_t$     as state equation,

where $\varepsilon_t \sim \mathcal{N}(0, \sigma_t^2)$ and $\eta_t \sim \mathcal{N}(0, Q_t)$ are independent of all other unknowns.

The components of state used in this paper are:

1. Local Linear trend

    The local linear trend model is a popular choice for modelling trends because it quickly adapts to local variation, which is desirable when making short term predictions, as such predictions often come with implausibly wide uncertainty intervals (Brodersen, 2015). Furthermore, Kay H. Brodersen (2015) explained that there is generalization of the local linear trend model where the slope exhibits stationarity instead of obeying a random walk, this model can be written as:

    $\mu_{t+1} = \mu_t + \delta_t + \eta_{\mu,t}$ ,

    $\delta_{t+1} = D + \rho(\delta_t - D) + \eta_{\delta,t}$ ,

    where the two components of $\eta$ are independent.

2. Seasonality

    The most frequently used model in the time domain is (Brodersen, 2015)

    $\gamma_{t+1} = -\sum_{s=0}^{S-2} \gamma_{t-s} + \eta_{\gamma,t}$ ,

    Where S represents the number of seasons and $\gamma_t$ denotes their joint contribution to the observed response $y_t$.

    **c) The Procedures of Predicting Tourism Demand Using Google Trends Data**

    First, related queries of international and domestic tourism demand were chosen by using Google Correlate and related queries provided by Google Trends. Based on ONS (2011) methods, their first, second, and third step were used in this paper. Then, the plot of Google Trends data in several current years were made by using R package gtrendsR, while the data was extracted manually from Google Trends tool. Simple average was conducted to the Google Trends data that already extracted.

The prediction of tourism demand in Indonesia based on Google Trends data was made separately for each of international and domestic tourism demand by using Bayesian Structural Time Series (BSTS) model. Then the error model was compared. Furthermore, the correlation of Google Trends data and official data was calculated.

## B.   Exploring Google Trends Data

Based on ONS (2011), the first step was determining three search term for all categories, such as "Bali", "wisata", "wisata di". Appendix 1 shows the trends of related keywords filtered by web, news, and image search from January 2004 up to December 2017. Word "Bali" represents the most popular tourism destination in Indonesia, "wisata" is "tourism", and "wisata di" is "tourism in" which has high correlation with word "wisata". Word "Bali" has the highest search hits compared to the other two words shown in Figure 18.

In Figure 19, word "wisata di" was replaced by word "Yogyakarta". Yogyakarta is also a famous tourism destination besides Bali. The search hits of word "Yogyakarta" is the lowest compared to word "Bali" and "wisata". In Figure 20, word "Bandung" was used to replace word "Yogyakarta". Bandung is also a famous tourism destination in Indonesia especially for people who live in area of Jakarta and surroundings. Word "Bandung" has higher search hits than word "wisata". In Figure 21, word "Bandung" was replaced by word "Flights to Indonesia" that has the lowest search hits.

The next filter is news and image search for all categories in January 2008 until December 2017. In news search, words "Bali" and "wisata" have more fluctuation than "wisata di" that can be seen in Figure 22. Meanwhile, in image search, words "Bali" and "wisata" were more stable than the trend in news search.

Figure 23 shows that the fluctuation occurs after 2014. In Figure 24, if the category was changed into travel category, the fluctuation for web search was equal to the filter of all categories. However, words "wisata" and "wisata di" were more stable in the filter of travel category than all categories. In Figure 25, if the filter was changed into news search in travel category, word "Bali" was more fluctuate than filtered by web and image search. Meanwhile, in image search, word "Bali" in travel category was more stable than the filter of all categories (Figure 26).

The next trial is by selecting one or more locations where people make searches. Then, comparison was conducted across these locations. Using "flights to Indonesia" as queries during January 2004 - December 2017, from all categories and web search, the comparison was conducted across Australia, Indonesia, and Singapore which is shown in Appendix 2. Figure 27 shows the very extreme fluctuations of search intensity happened for queries "flights to Indonesia" in Indonesia and Singapore. The very extreme fluctuations happened in Indonesia only during 2004-2010, while in Singapore it happened since 2004-2017. If the category was changed into travel category, it reduces the fluctuations happened, but in Singapore the fluctuations still happened in all of observation periods (Figure 28). The fluctuations in Singapore may happened because of the distance between Indonesia and Singapore is very close. The language of queries also influences the result of search intensity provided by Google Trends, for instance, word "wisata" as the Indonesian language of "tourism" will be resulted in the bigger number of search intensity in Indonesia compared to search intensity in Australia and Singapore.

The last step of trial based on ONS (2011) was by selecting one or more-time range, then compares the interest in a search term in these periods. In this paper, there are three periods that were compared that can be seen in Appendix 3 January

2004 - December 2017, January 2008 - December 2017, and January 2016 - December 2017. In January 2004 - December 2017, the search intensity tends to increase and during 2005 was the most fluctuated search intensity (Figure 33). If the time range was shorted to become January 2008 - December 2017, the fluctuations can be seen clearer than the previous time range (Figure 34), as well as in January 2016 - December 2017 when the fluctuations can be seen very clearly and the positive trends cannot be seen as the previous two times range (Figure 35).

Exploring Google Trends tool reveals its strengths and weaknesses. Google Trends can provide a nowcasting figure of several indicators, it can fill the gaps in providing tourism statistics which has data lag in the data release schedule. In order to use Google Trends data, firstly, user has to be careful in choosing the related queries to approach an indicator. It really matters that each query needs special treatment, so that user needs several trials.

Using Google Trends via its website, it cannot provide the comparison of search intensity of a query among countries, R package gtrendsR can be used in order to solve this problem. However, both of Google Trends website and gtrendsR cannot be used to compare the search intensity among type of search and categories in one time when running the R codes. Furthermore, list of categories in Google Trends website are limited compared to list of categories in gtrendsR. Google Trends website only provides 26 list of categories including all categories as an option, while gtrendsR provides 1,782 categories. In addition, if user wants to see the search intensity in a region, user cannot compare it all in one time, but by clicking the region one by one. Country and sub-country or region and city codes are provided in gtrendsR so that user can compare it all in one time when running the R codes.

In case of choosing one or more-time range then compare the search intensity of a query, it has a weakness when using Google Trends website and gtrendsR in one time, it cannot be used to compare more than one-time range. In the same time range and type of search, but different category, it can be resulted the differences in patterns and fluctuations. In looking at the search intensity of a query in country which has different language with query that user used, it will impact on the low number of search intensity resulted. For instance, using word "wisata" with Australia as the location of search will be resulted in low number of search intensity. It will be different if the word "tourism" was used in Australia as the location. It is because word "wisata" is in Indonesian language which is used in English speaking country.

## C. Predicting Tourism Demand in Indonesia Using Google Trends Data

### 1. International Tourism Demand in Indonesia Based on Google Trends Data

Figure 1 shows several queries related to people who search the information for travelling to Indonesia, words "flights to Indonesia", "Indonesia time", "Indonesia travel", "Indonesia visa", and "travel to Indonesia". The pattern of four queries are nearly the same, except for word "Indonesia time" which has trends since 2016. The highest search hits are words "Indonesia time", followed by words "Indonesia visa" and "Indonesia travel". The pattern of some words (not only in Figure 1) have trends since 2015. It was may happen because of since 2015 a program from Ministry of Tourism called "Wonderful Indonesia" was promoted massively in all over the world. In 2015, there was a series of events organized to celebrate the 60th Asian-African Conference which held in Jakarta and Bandung.

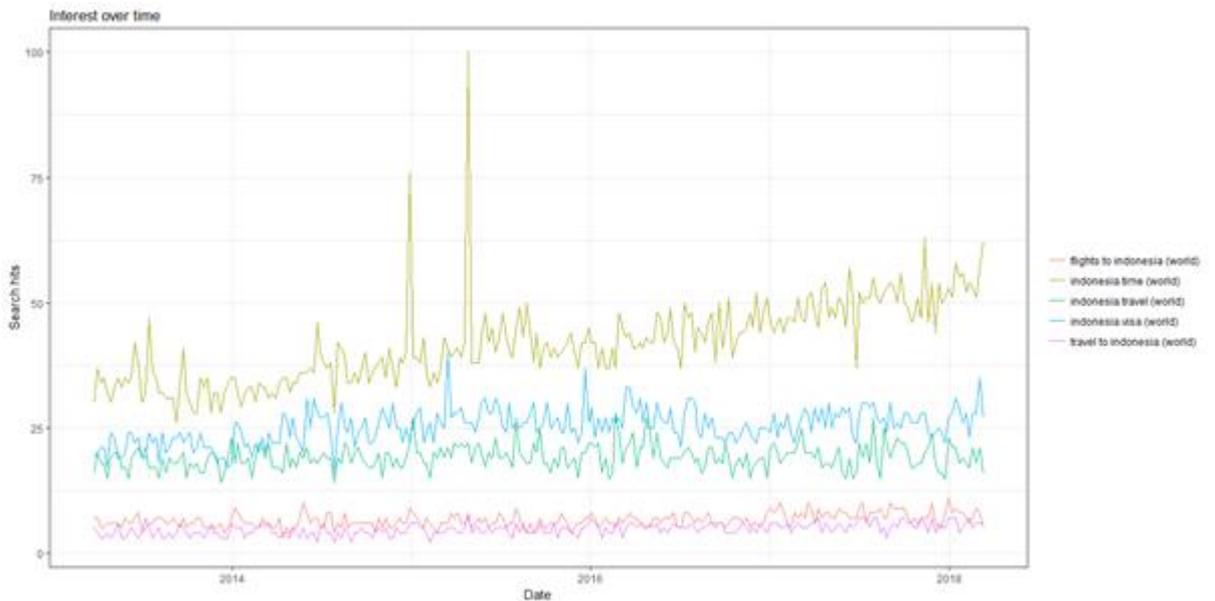

Figure 1. Plot of Keyword (Flights to Indonesia, Indonesia Time, Indonesia Travel, Indonesia Visa, Travel to Indonesia)

In order to predict tourism demand in Indonesia, the prediction for tourism demand from international and domestic side were separated. In this case, queries used to predict tourism demand in Indonesia from international side were flights to Indonesia, Indonesia time, Indonesia travel, Indonesia visa, and travel to Indonesia. Then simple average was applied for these queries.

Figure 2 shows plot of Bayesian Structural Time Series (BSTS) model of international tourism demand in Indonesia by using Google Trends data in 2004 – 2017. The BSTS model used includes local linear trends and trig as state specification. In 2004 to 2006 the prediction shows a tendency to decrease, but after that it tends to increase up to 2017. This prediction shows that the 2005 Bali bombing cause a decrement of tourism demand in Indonesia. The prediction also shows several small fluctuations during the period of observation.

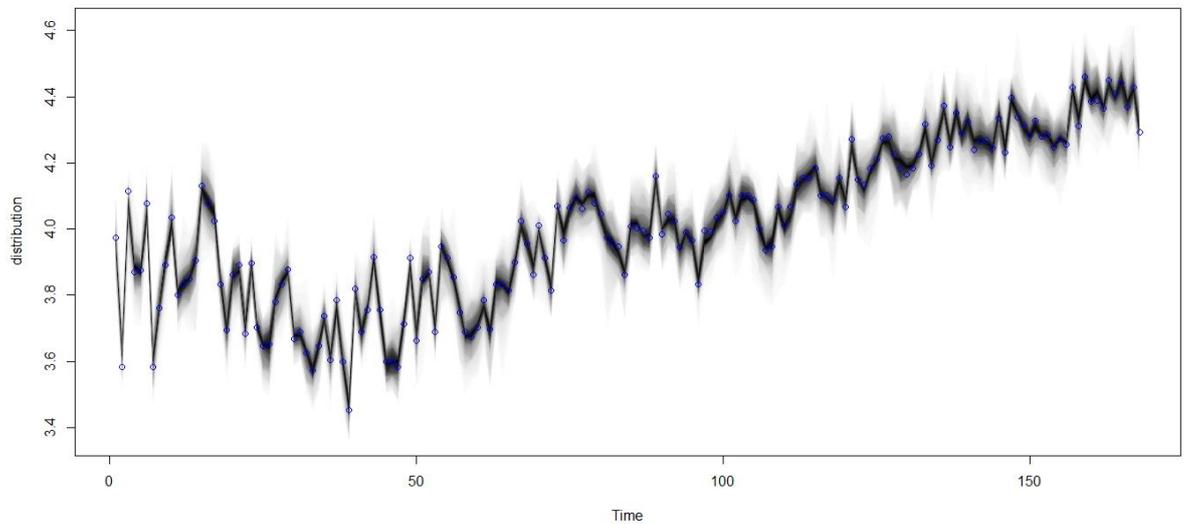

Figure 2. Plot of BSTS Model of International Tourism Demand in Indonesia 2004 – 2017 Based on Google Trends Data

Figure 3 shows the comparison of error from BSTS model which includes trend compared to model with trend and seasonal. The error shows similar pattern during January 2004 to December 2017. The differences between two error models are relatively small, but after ten years (after 2014), the differences are getting bigger. The small differences of error between model with trend and model with seasonal shows that to predict the tourism demand from international side, it can be used model with trend or model with seasonal.

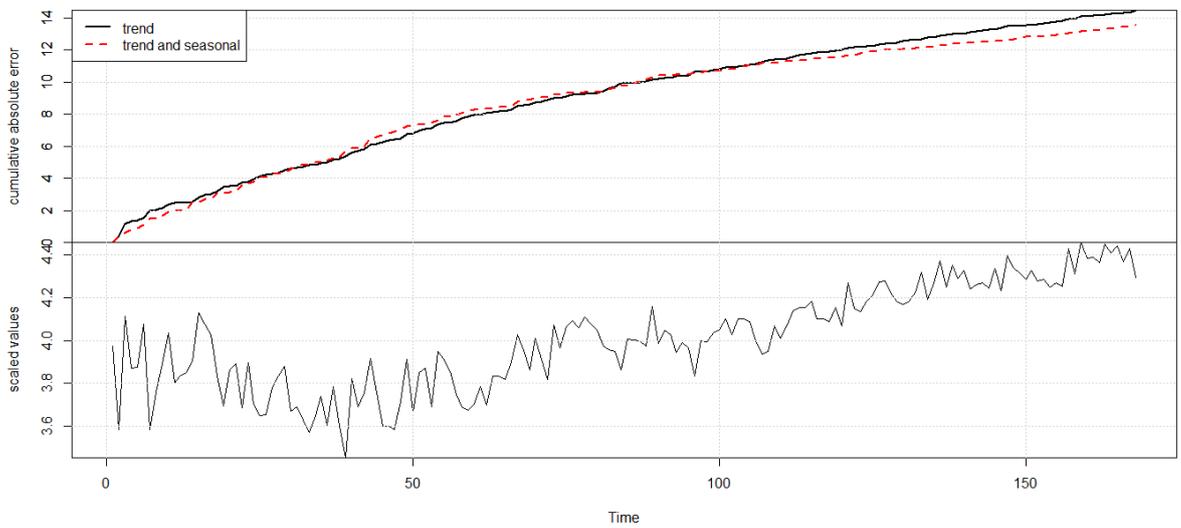

Figure 3. BSTS Model of International Tourism Demand in Indonesia 2004 – 2017 Based on Google Trends Data: Comparison of Error from Model with Trend and Model with Trend and Seasonal

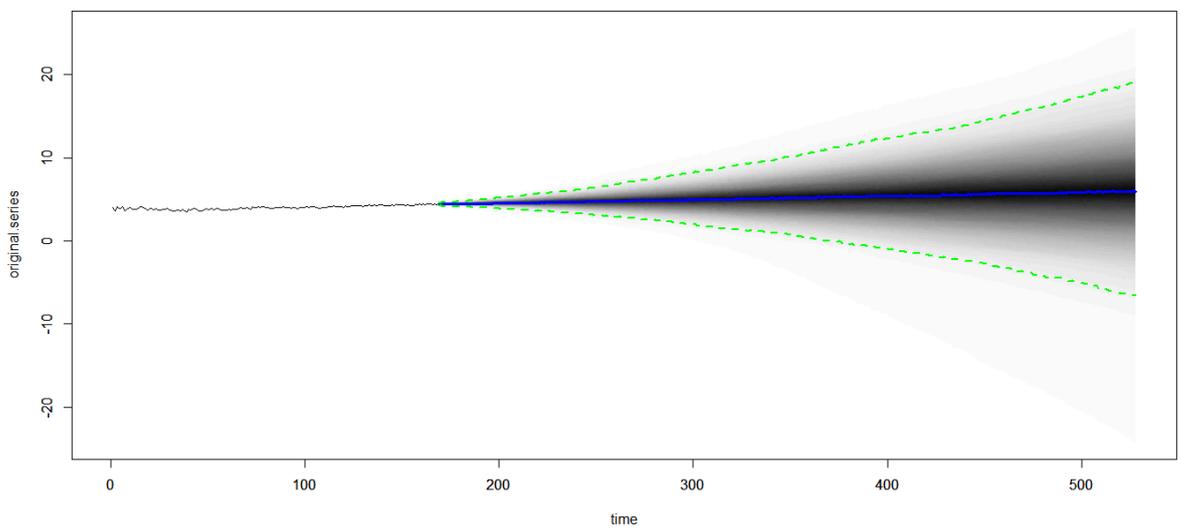

Figure 4. BSTS Model of International Tourism Demand in Indonesia Based on Google Trends Data: Local Linear Trend Prediction with Confidence Interval

The prediction of tourism demand from international visitor based on Google Trends data can be seen in Figure 4 below. Figure 4 shows the prediction with confidence interval (green line). This prediction was resulted from the BSTS model with local linear trend. Moreover, Figure 5 shows the prediction of international tourism demand from Google Trends data by using semi-local linear trend. The confidence interval of this prediction is narrower than using local linear trends.

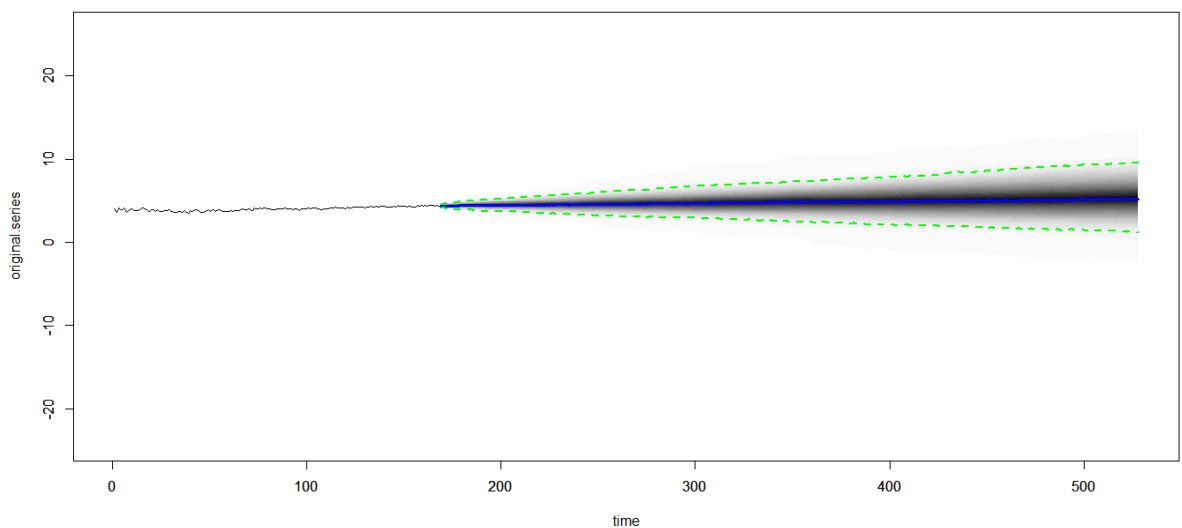

Figure 5. BSTS Model of International Tourism Demand in Indonesia Based on Google Trends Data: Semi-Local Linear Trend with Confidence Interval

2. **Domestic Tourism Demand in Indonesia Based on Google Trends Data**

Firstly, queries related to tourism demand of domestic visitor were created using Google Correlate and related queries shown in the result of Google Trends. Figure 6 shows the plot of "tempat wisata" or "tourism place", "wisata" or "tourism", and "wisata di" or "tourism in" resulted from Google Trends data. These three queries show similar pattern during the period of observation, although they have different search hits. Query "wisata" has the highest search hits among the other two queries. The search hits of these three queries tend to increase during the period of observation.

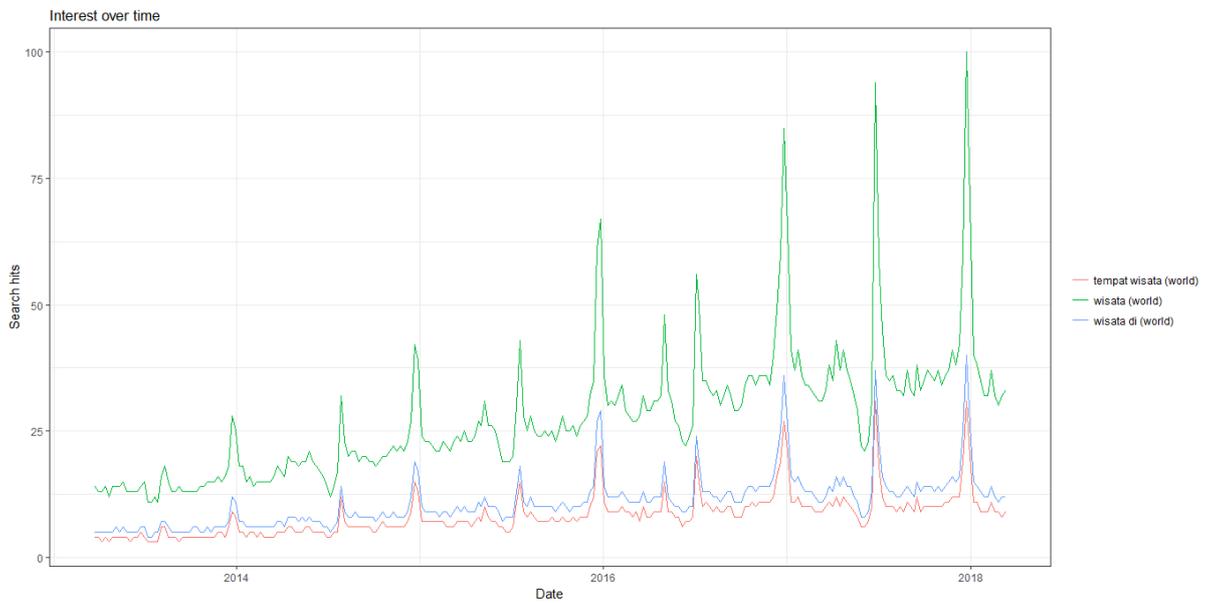

Figure 6. Plot of Words "Tempat Wisata", "Wisata", and "Wisata di"

Figure 7 shows plot of BSTS model of domestic tourism demand in Indonesia during January 2004 to December 2016. This BSTS model used local linear trend and trig for state specification. Overall, the model shows a tendency of increasing although in several points it shows a decreasing.

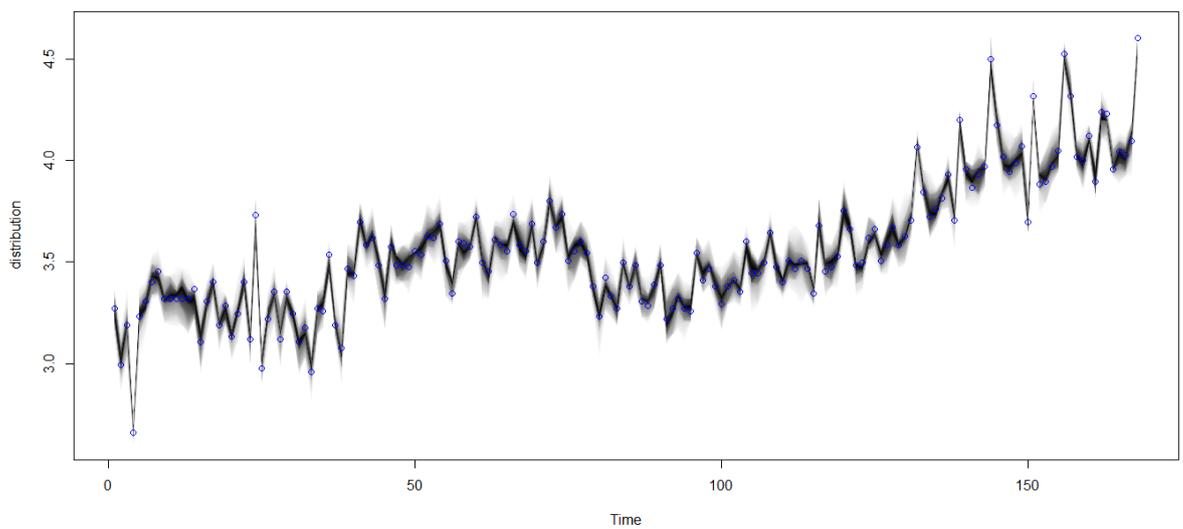

Figure 7. BSTS Model of Domestic Tourism Demand in Indonesia 2004 – 2017 Based on Google Trends Data

The differences between error model of model with trend and model with trend and seasonal can be seen in Figure 8. Figure 8 shows that the differences between two models are small in number of cumulative absolute error. The error of model with trend and seasonal is bigger than error of model with trend only without seasonal. However, Figure 8 shows that the error of two model has similar pattern during the period of observation.

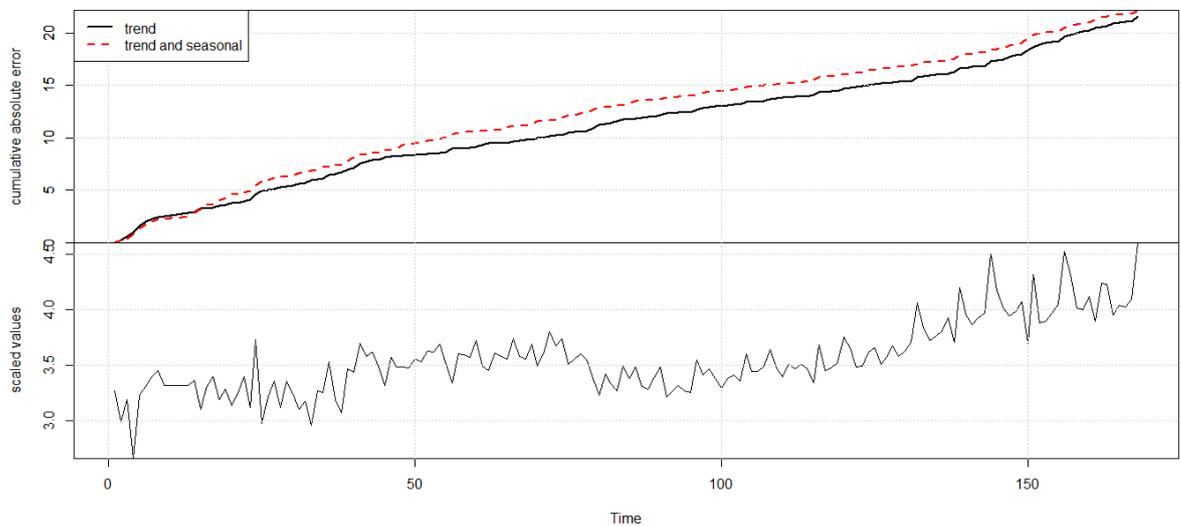

Figure 8. BSTS Model of Domestic Tourism Demand in Indonesia 2004 – 2017: Comparison of Error from Model with Trend and Model with Trend and Seasonal

The confidence interval of the BSTS model for the prediction of domestic tourism demand in Indonesia is shown in Figure 9. Figure 9 shows the prediction based on Google Trends data by including local linear trend. The confidence interval has a wide gap between the lower and upper bound represented by the green line. Prediction model with semi-local linear trend has a narrower gap of confidence interval than prediction model with local linear trend, it is shown by Figure 10.

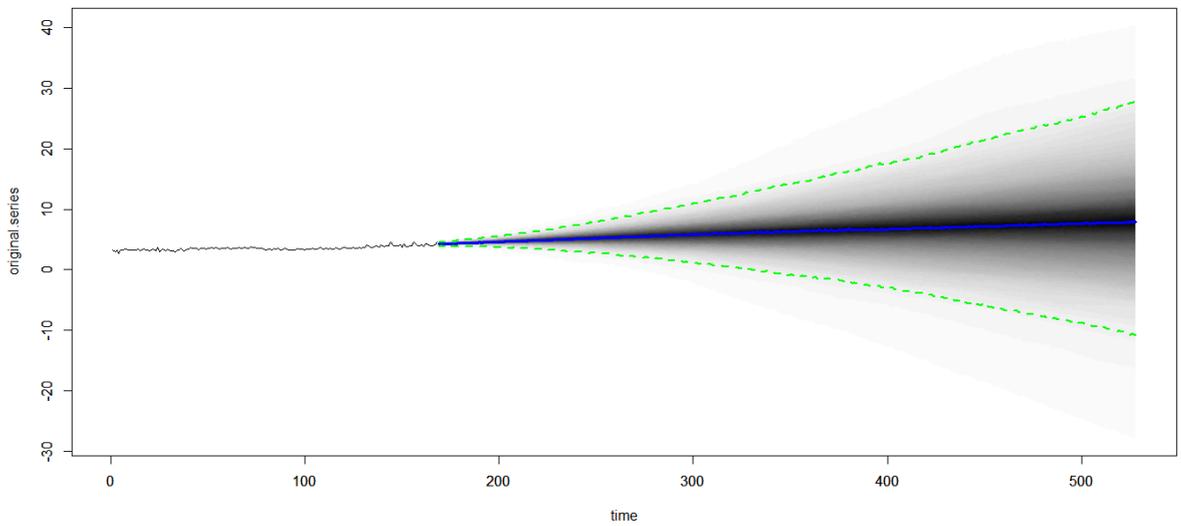

Figure 9. BSTS Model of Domestic Tourism Demand in Indonesia Based on Google Trends Data: Local Linear Trend Prediction with Confidence Interval

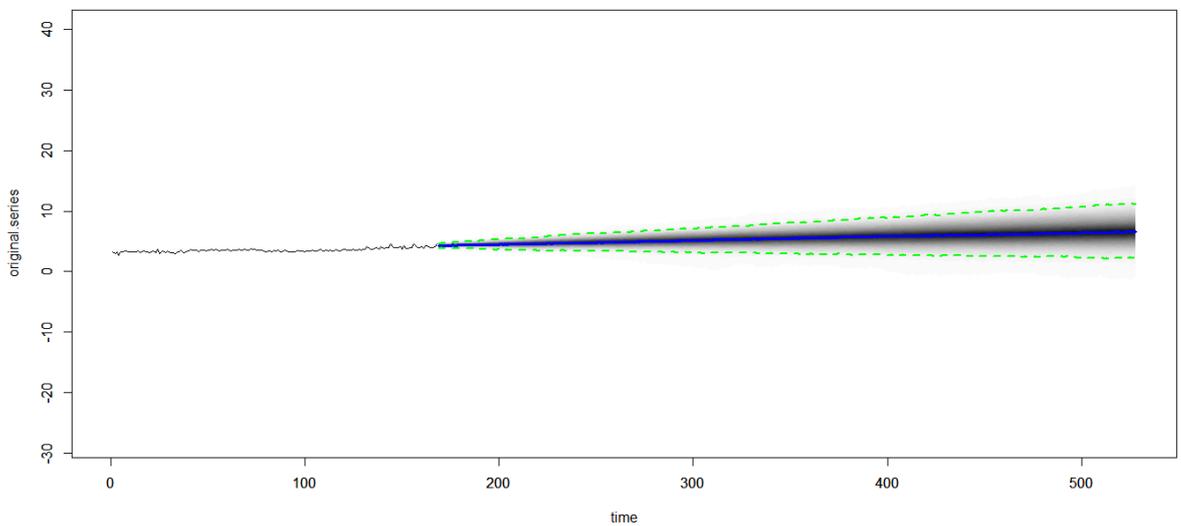

Figure 10. BSTS Model of Domestic Tourism Demand in Indonesia Based on Google Trends Data: Semi-Local Linear Trend Prediction with Confidence Interval

## D. Comparison between Google Trends Data and Official Data

In order to investigate whether Google Trends data can show the pattern of tourism demand in Indonesia or not, the next step was by comparing the error model resulted from the prediction of data from Google Trends and official data from BPS-Statistics Indonesia. Figure 11 shows the comparison among Model 1, Model 2, and BPS model in predicting international tourism demand in Indonesia. Model 1 resulted from Google Trends data with local prediction, Model 2 resulted from Google Trends data with semi-local prediction, while BPS model resulted from official data released by BPS-Statistics Indonesia. Figure 11 also shows that the three models have similar pattern during the period of observation.

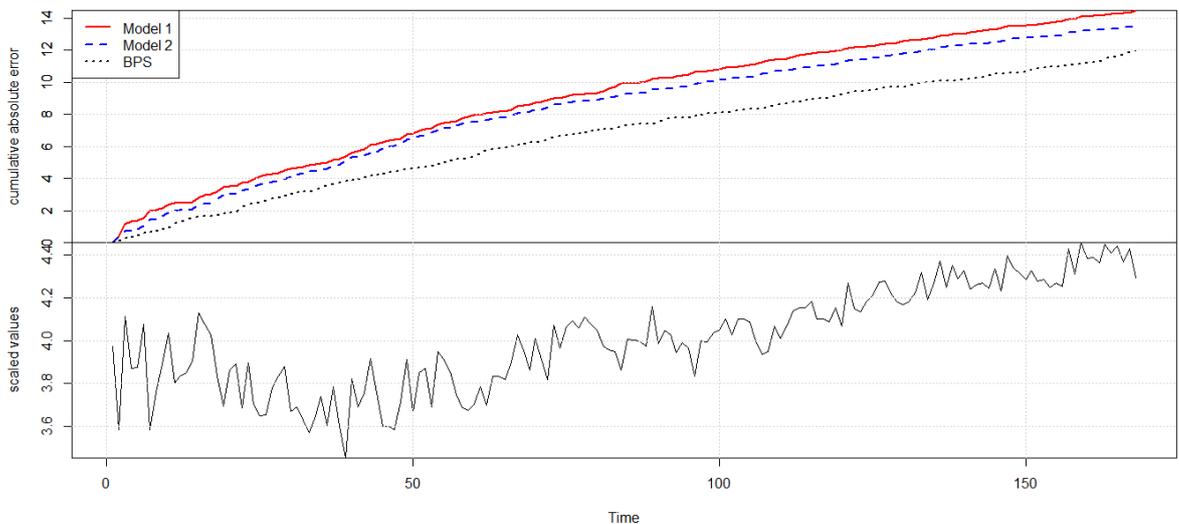

Figure 11. Prediction of International Tourism Demand: Comparison of Error from Model 1 (Google Trends Data with Local Prediction), Model 2 (Google Trends Data with Semi-Local Prediction), and Model 3 (Official Data from BPS)

Table 1. Summary of Model 1, Model 2, and BPS Model

| Model | Residual.sd | Prediction.sd | R Square |
|---|---|---|---|
| (1) | (2) | (3) | (4) |
| Model with Local Trends (GT data) | 0.1476276 | 0.1748606 | 0.7856612 |
| Model with Semi-Local Trends (GT data) | 0.1241448 | 0.1669883 | 0.8484268 |
| Model of Official Data | 0.07350634 | 0.08964715 | 0.9500723 |

Table 1 is a summary of BSTS model for predicting international tourism demand. The smallest posterior mean of the residual standard deviation parameter (residual.sd) is model of official data, followed by model with semi-local trends (Google Trends data). In addition, the smallest standard deviation of the one-step-ahead prediction errors (prediction.sd) is also model of official data, followed by model with semi-local trends (Google Trends data). R-square of official data model is 0.9500723 which means that the fit explains 95.01 percent of the total variation in the data about the average. Meanwhile, R-square of Model 1 (with local trend) shows that the fit can explain 78.57 percent of the total variation in the data about the average, and the fit of Model 2 (with semi-local trend) can explain 84.84 percent of the total variation in the data about the average.

Figure 12 shows the comparison among Domestic Visitor GT 1, Domestic Visitor GT 2, and BPS model in predicting domestic tourism demand in Indonesia. Model of Domestic Visitor GT 1 resulted from Google Trends data with local prediction, Model of Domestic Visitor GT 2 resulted from Google Trends data with semi-local prediction, while BPS model resulted from official data released by BPS-Statistics Indonesia. Figure 15 also shows that models resulted from Google Trends data have different pattern with model resulted from official data.

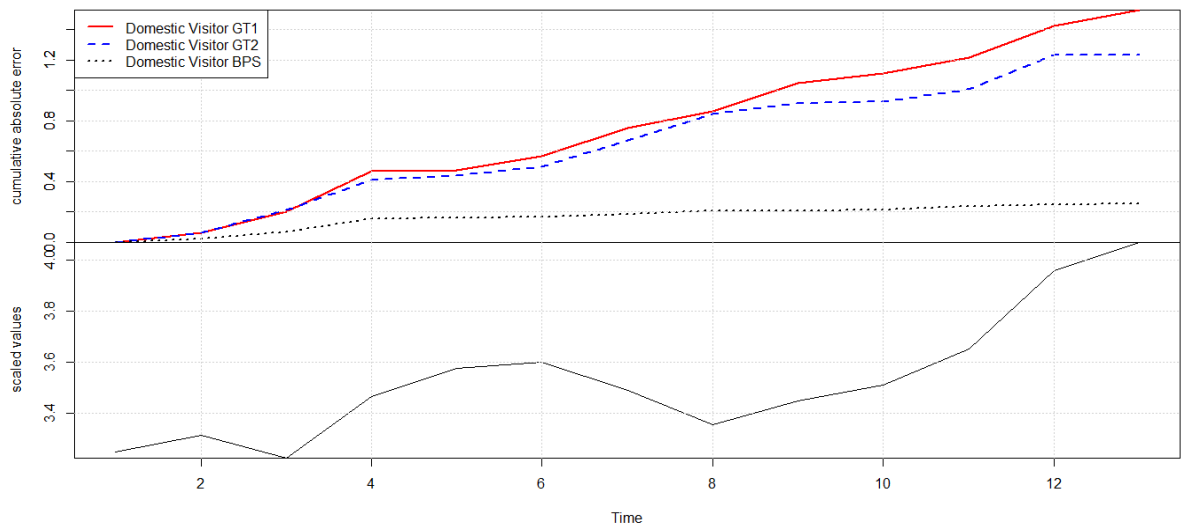

Figure 12. Prediction of Domestic Tourism Demand: Comparison of Error from Model 1 (Google Trends Data with Local Prediction), Model 2 (Google Trends Data with Semi-Local Prediction), and Model 3 (Official Data from BPS)

Table 2. Summary of Model for Predicting Domestic Tourism Demand in Indonesia

| Model | Residual.sd | Prediction.sd | R Square |
|---|---|---|---|
| (1) | (2) | (3) | (4) |
| Model with Local Trends (GT data) | 0.06445756 | 0.143981 | 0.9337223 |
| Model with Semi-Local Trends (GT data) | 0.01936385 | 0.03165525 | 0.9566947 |
| Model of Official Data | 0.04744663 | 0.1272231 | 0.9640888 |

Table 2 is a summary of BSTS model for predicting domestic tourism demand. The smallest posterior mean of the residual standard deviation parameter (residual.sd) is model with semi-local trends (Google Trends data), followed by of official data model. In addition, the smallest standard deviation of the one-step-ahead prediction errors (prediction.sd) is also model with semi-local trends (Google Trends data), followed by model of official data. R-square of official data model is 0.9640888 which means that the fit explains 96.41 percent of the total variation in the data about the average. Meanwhile, R-square of model with local trend shows that the fit can explain 93.37 percent of the total variation in the data about the average, and the fit of model with semi-local trend can explain 95.67 percent of the total variation in the data about the average.

Table 3. Correlation between Google Trends Data and Official Data

| Data | Correlation |
|---|---|
| (1) | (2) |
| International Tourism Demand | 0.8730973 |
| Domestic Tourism Demand | 0.772844 |

The correlation between Google Trends data and official data are good for international and domestic tourism demand data, it is shown in Table 3. International tourism demand from Google Trends data has 0.87 correlation value with official data. Meanwhile, the correlation between domestic tourism demand resulted from Google Trends data and official data was accounted for 0.77. It indicates that Google Trends data can be used as complement of official data in order to see the pattern of tourism demand in Indonesia.

## E.   Impact of Disaster on Google Trends Data: Case Study in Bali Province

Google Trends data also provided the break down by origin of search hits (region and city). In Bali Province, the highest search hits were from Central Java, Yogyakarta, and, East Java Province which is shown in Figure 13. Meanwhile, Figure 14 shows that Jembrana, Mengwi, and Klungkung as city of origin with the highest search hits within Bali Province.

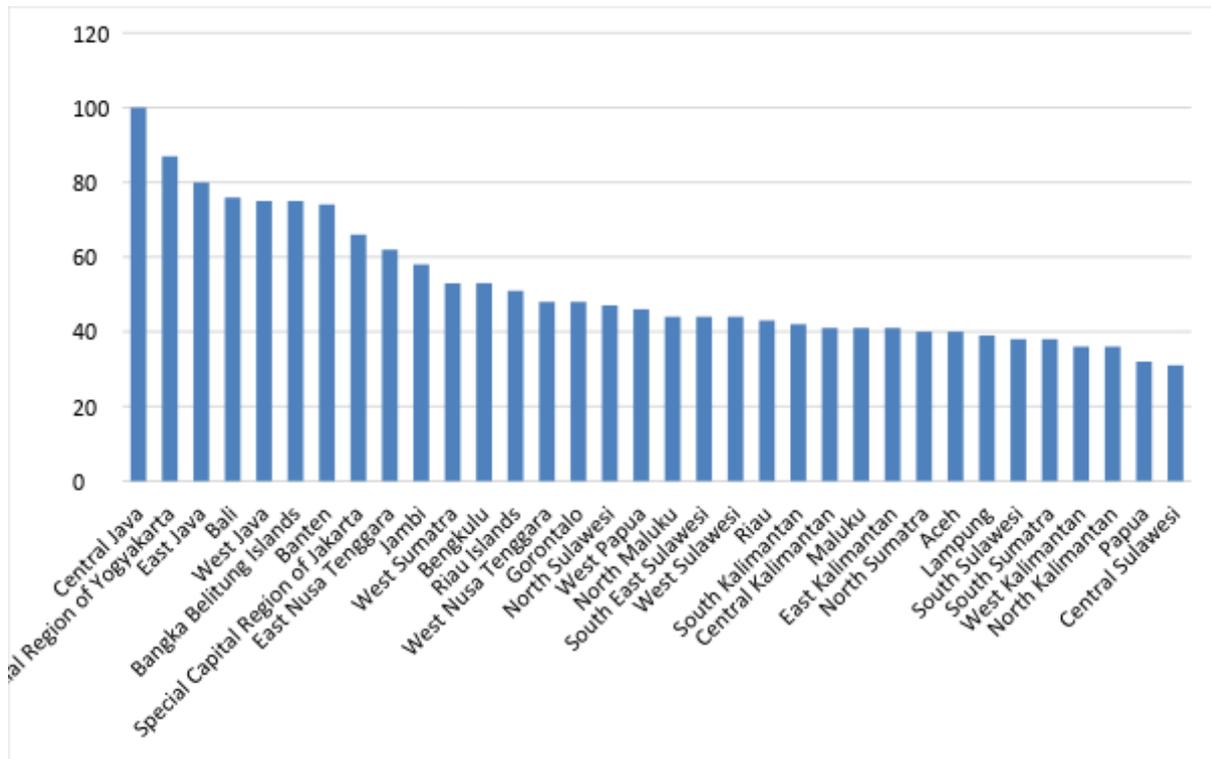

Figure 13. Search Hits of "Wisata" or "Tourism" in Bali Province by Region of Origin in Indonesia Based on Google Trends Data

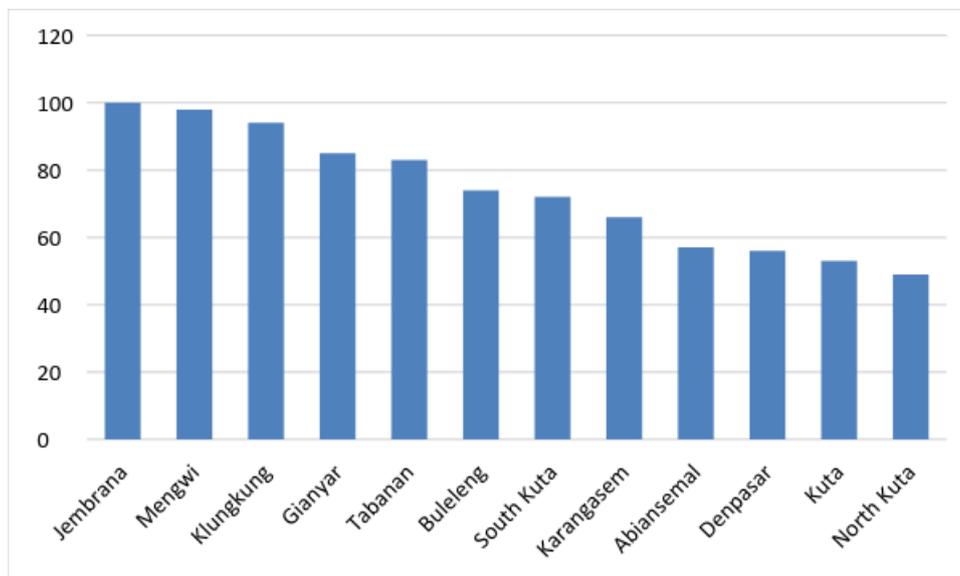

Figure 14: Search Hits of "Wisata" or "Tourism" in Bali Province by City of Origin in Bali Based on Google Trends Data

Disaster has an impact on number of visitor in an area, as well as disaster happened in Bali Province, such as volcanic eruption and bombings. Figure 15 shows the country of origin who searched about "Bali Volcano". The current volcanic eruption in Bali happened in October – December 2017. Figure 16 shows the decreasing of search hits for "Hotel in Bali" and the increasing of search hits for "Bali Volcano" in October and December 2017.

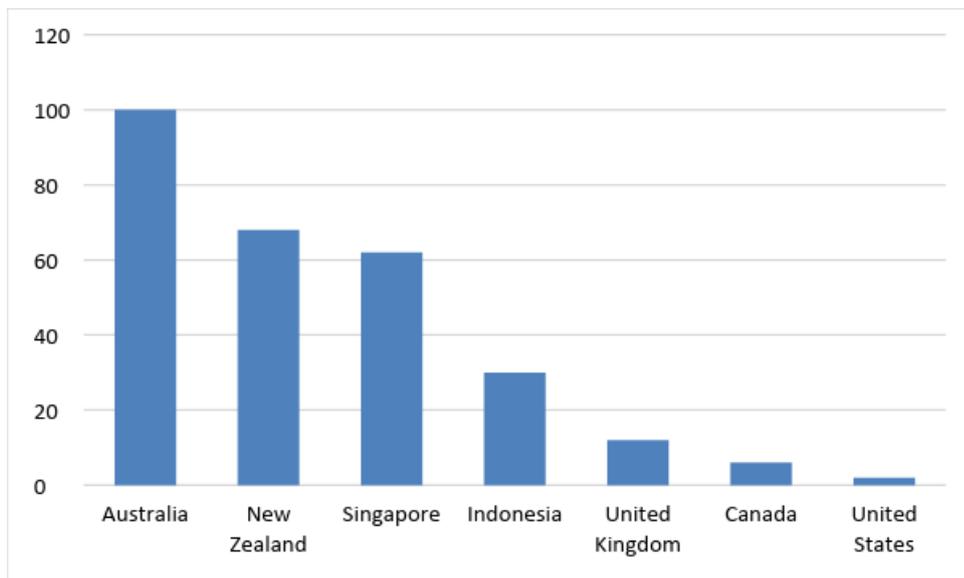

Figure 15. Search Hits of "Bali Volcano" by Country of Origin 2004 – 2017

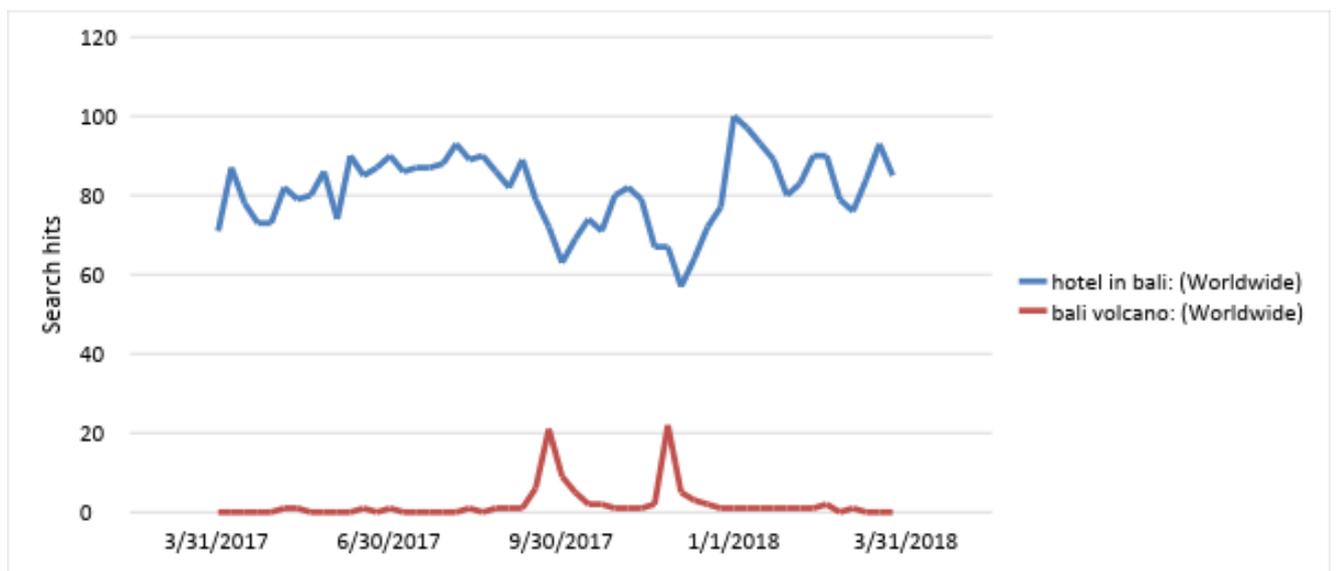

Figure 16. Search Hits of "Hotel in Bali" and "Bali Volcano"

The causal impact of volcanic eruption in Bali can be inferred using CausalImpact package on R found by Brodersen et. al. (2015). In relative terms, it can be inferred that the number of international visitor arrival showed a decrease of -2 percent. The 95% interval of this percentage is [-4%, +0%]. Although it may look as though the intervention has exerted a negative effect on the number of international visitor arrival when considering the volcanic eruption period as a whole, this effect is not statistically significant, and so cannot be meaningfully interpreted. The apparent effect could be the result of random fluctuations that are unrelated to the volcanic eruption in Bali. It can be the case when the period of volcanic eruption in Bali is too short to distinguish the signal from the noise. However, the probability of obtaining this effect by chance is very small (Bayesian one-sided tail-area probability $p = 0.037$). This means the causal effect can be considered statistically significant.

The Bali bombings happened in 2002 and 2005, while Google Trends provides data only from 2004. However the impact of the 2002 Bali bombings still can be seen in Figure 17. In 2005, there was a sharp increasing of search hits for "Bali bombings". Search hits for "hotel in Bali" started to increase in 2007 based on Figure 17.

In addition, if the causal impact of Bali bombing also be inferred using CausalImpact package, the number of international visitor showed a decrease of -5 percent. The 95% interval of this percentage is [-6%, -3%]. This means that the negative effect observed during the Bali bombing period is statistically significant. The summary report of this causal impact shows that the probability of obtaining this effect by chance is very small (Bayesian one-sided tail-area probability $p = 0.001$). This means the causal effect can be considered statistically significant.

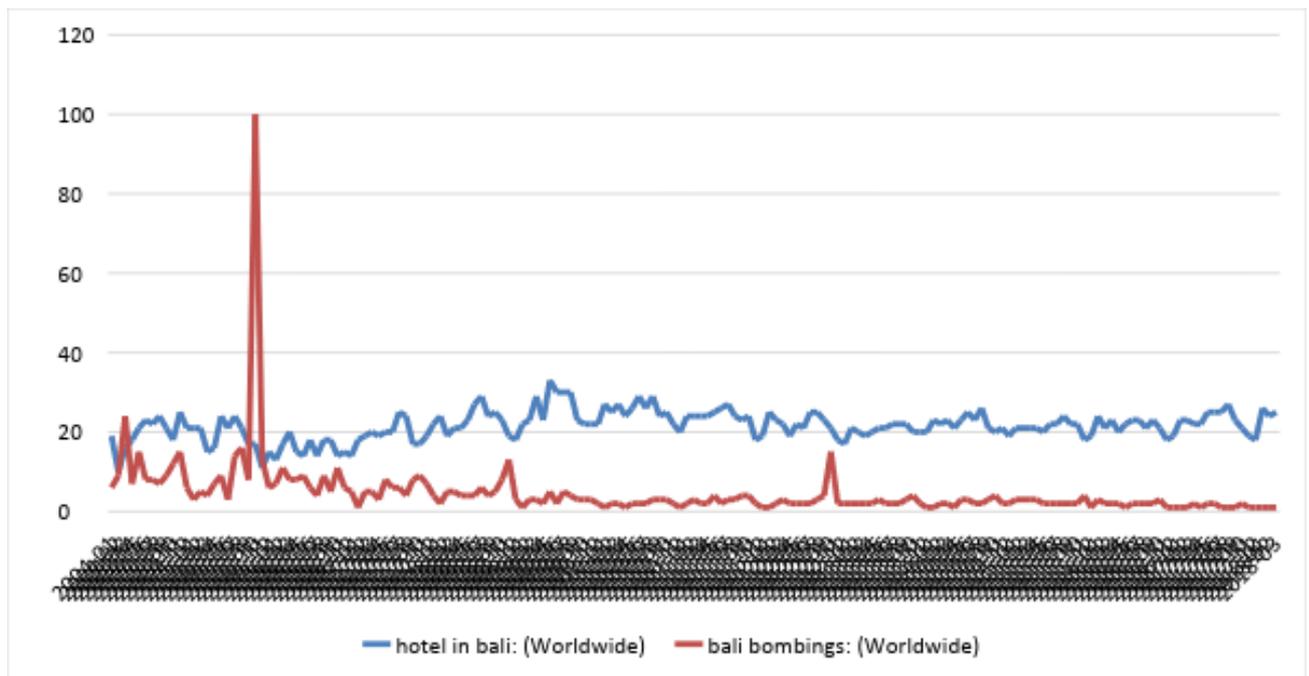

Figure 17. Search Hits of "Hotel in Bali" and "Bali Bombings"

## III. Conclusion

The result of prediction from BSTS model shows that Google Trends can show the pattern of tourism demand in Indonesia because it has similar pattern especially for international tourism demand. It suggests that the queries for predicting domestic tourism demand from Google Trends data need an additional queries then compare it to find out the most representative queries. The result of correlation shows that Google Trends data has good correlation with official data. Disaster has an impact on the decreasing of search hits for tourism related queries on Google Trends data.

Google Trends can provide a nowcasting figure of several indicators, it can fill the gaps in providing tourism statistics which has data lag in the data release schedule. It is in order to achieve the 2018 Indonesian Government Work Plans, to fulfil the criteria in accuracy, relevant, actual, timeliness, accessibility, and coherent to support the planning and evidence-based policy. Then, it is also important to accelerate the achieving of the 2030 Agenda of SDGs. However, Google Trends have several weaknesses that need to be considered. It really matters that each query needs special treatment, so that user needs several trials for predicting an indicator up to getting the standard procedures.

This is the first paper which addressing Google Trends data for predicting tourism demand in Indonesia, including analysis of tourism demand in regencies/cities and considering the effect of disaster. Future study is needed to enrich the knowledge of using Google Trends data and to get the standard procedures for estimating various indicators in different countries in the world.

Therefore, apart from its strength and weakness, Google Trends data can be used as complement and new approach of official data in order to see the pattern of tourism demand in Indonesia. It can be an early warning for policy maker in achieving their target of tourists.

# Appendices

Appendix 1

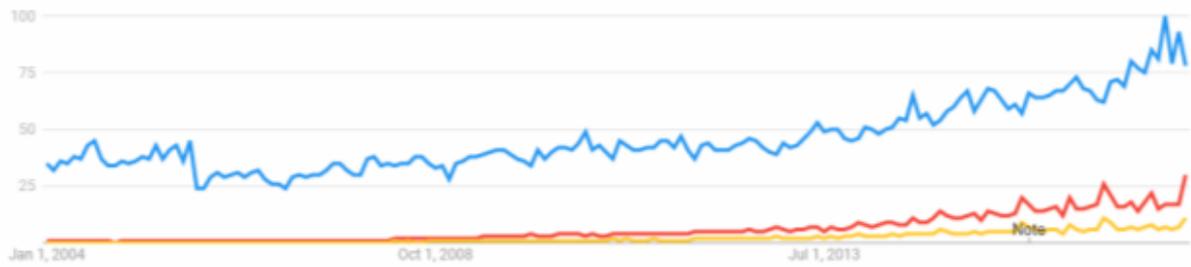

Figure 18. Plot of Keywords (Bali, Wisata, Wisata di) for All Categories in Web Search

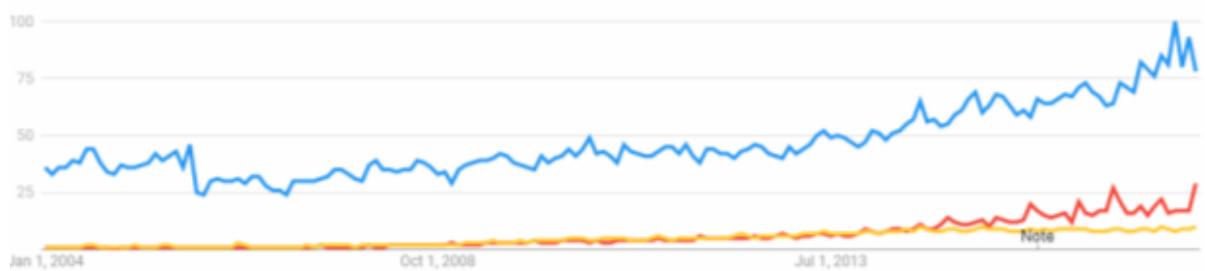

Figure 19. Plot of Keywords (Bali, Wisata, Yogyakarta) for All Categories in Web Search

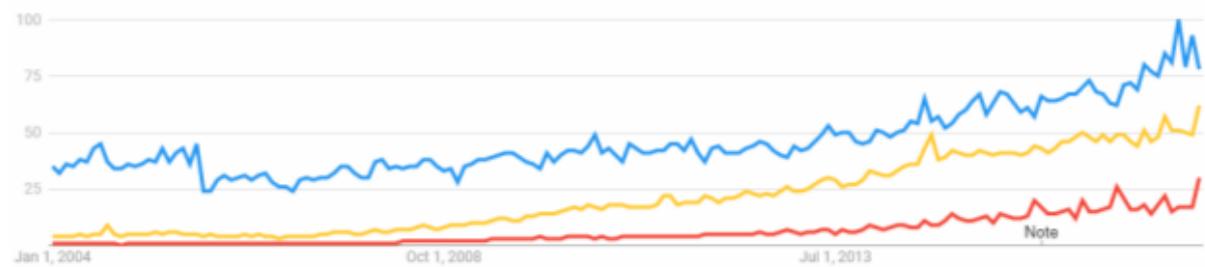

Figure 20. Plot of Keywords (Bali, Bandung, Wisata) for All Categories in Web Search

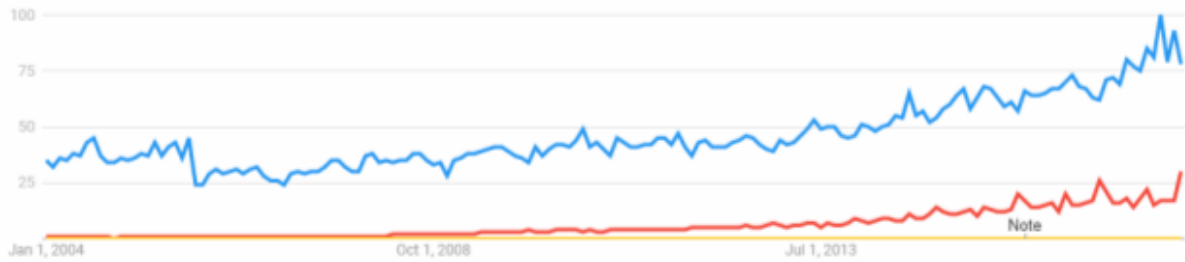

Figure 21. Plot of Keywords (Bali, Wisata, Flights to Indonesia) for All Categories in Web Search

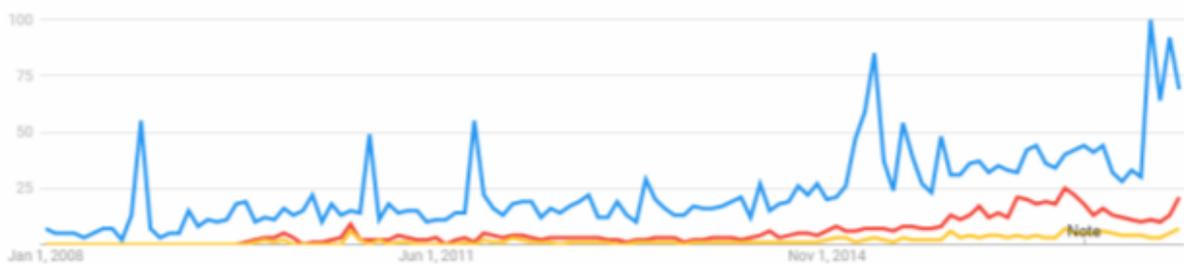

Figure 22. Plot of Keywords (Bali, Wisata, Wisata di) for All Categories in News Search

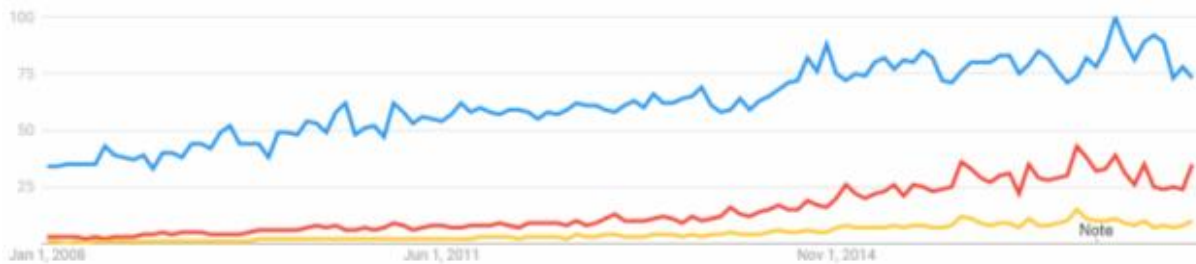

Figure 23. Plot of Keywords (Bali, Wisata, Wisata di) for All Categories in Image Search

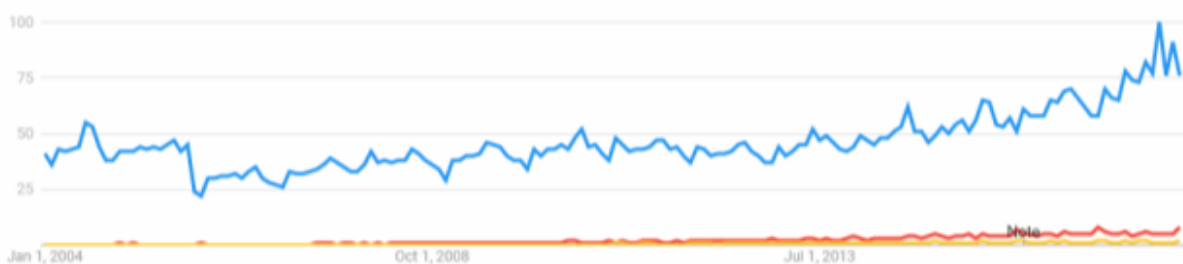

Figure 24. Plot of Keywords (Bali, Wisata, Wisata di) for Travel Category in Web Search

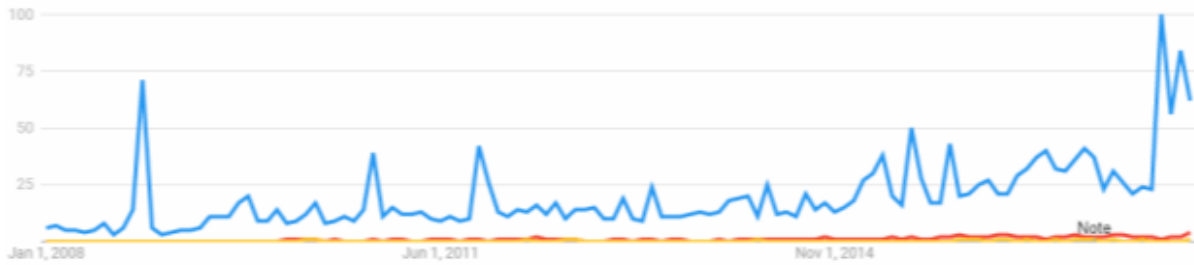

Figure 25. Plot of Keywords (Bali, Wisata, Wisata di) for Travel Category in News Search

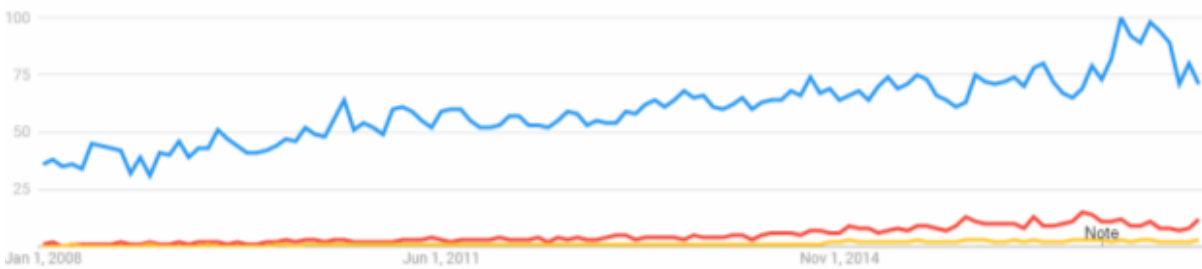

Figure 26. Plot of Keywords (Bali, Wisata, Wisata di) for Travel Category in Image Search

Appendix 2

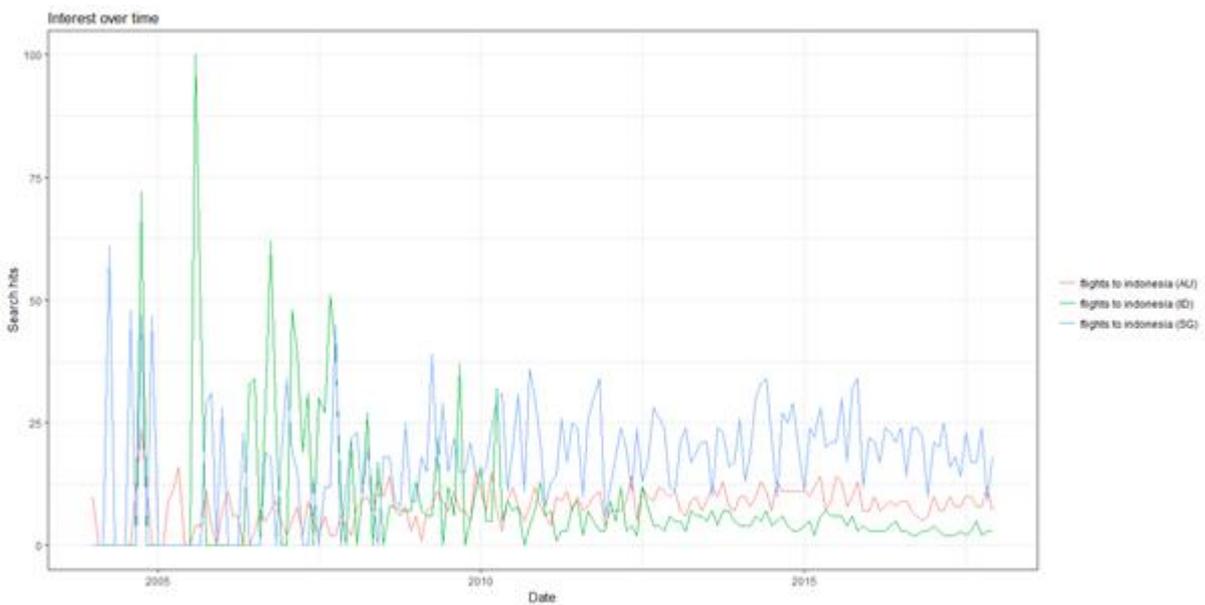

Figure 27. Plot of "flights to Indonesia" in Australia (AU), Indonesia (ID), and Singapore (SG) for All Categories in Web Search, 2004-2017

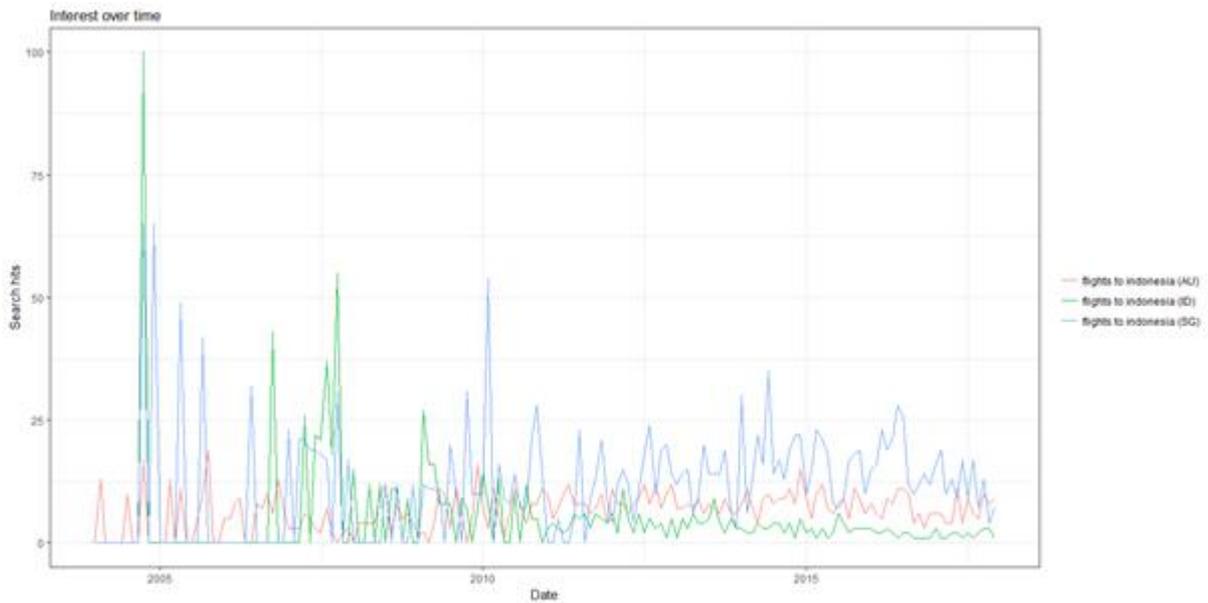

Figure 28. Plot of "flights to Indonesia" in Australia (AU), Indonesia (ID), and Singapore (SG) for Travel Category in Web Search, 2004-2017

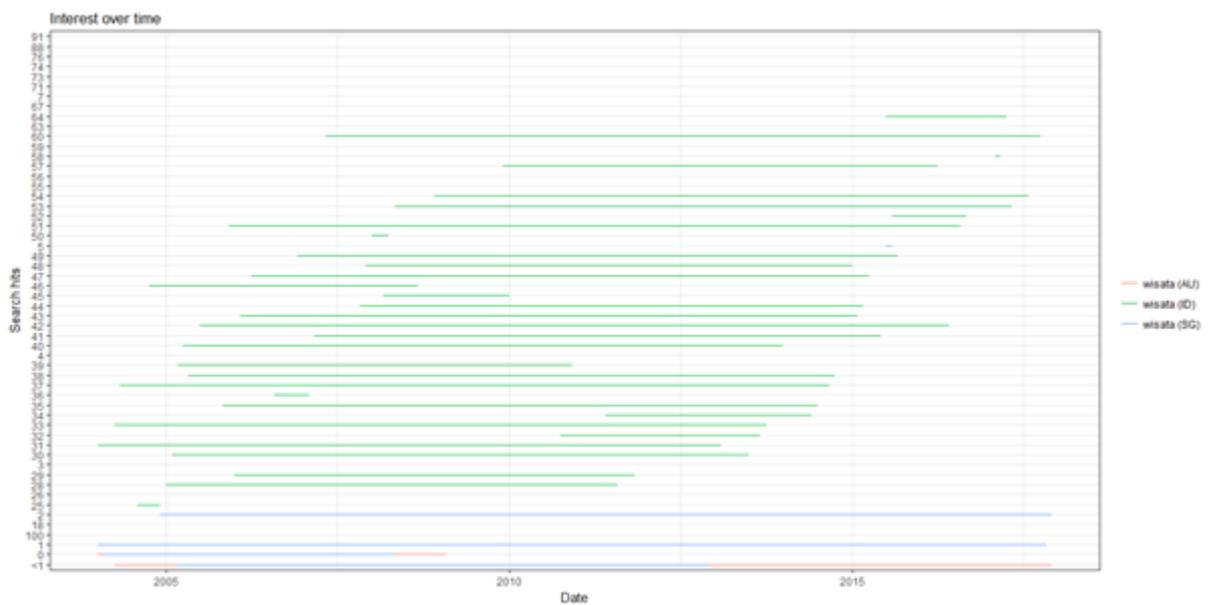

Figure 29. Plot of "wisata" in Australia (AU), Indonesia (ID), and Singapore (SG) for All Categories in Web Search, 2004-2017

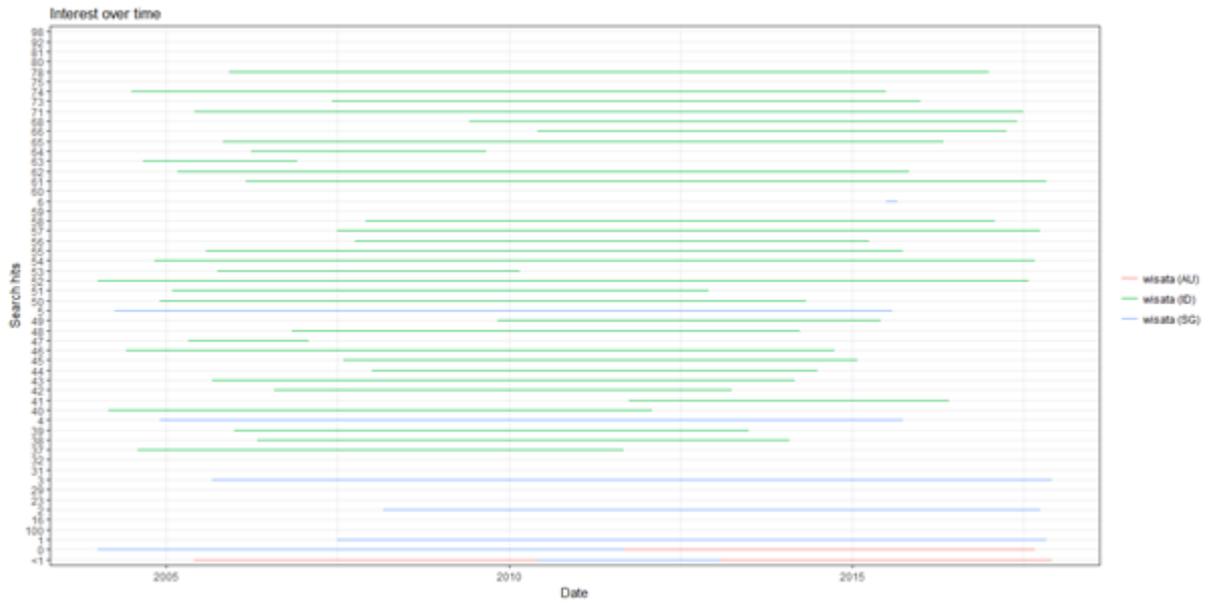

Figure 30. Plot of "wisata" in Australia (AU), Indonesia (ID), and Singapore (SG) for Travel Category in Web Search, 2004-2017

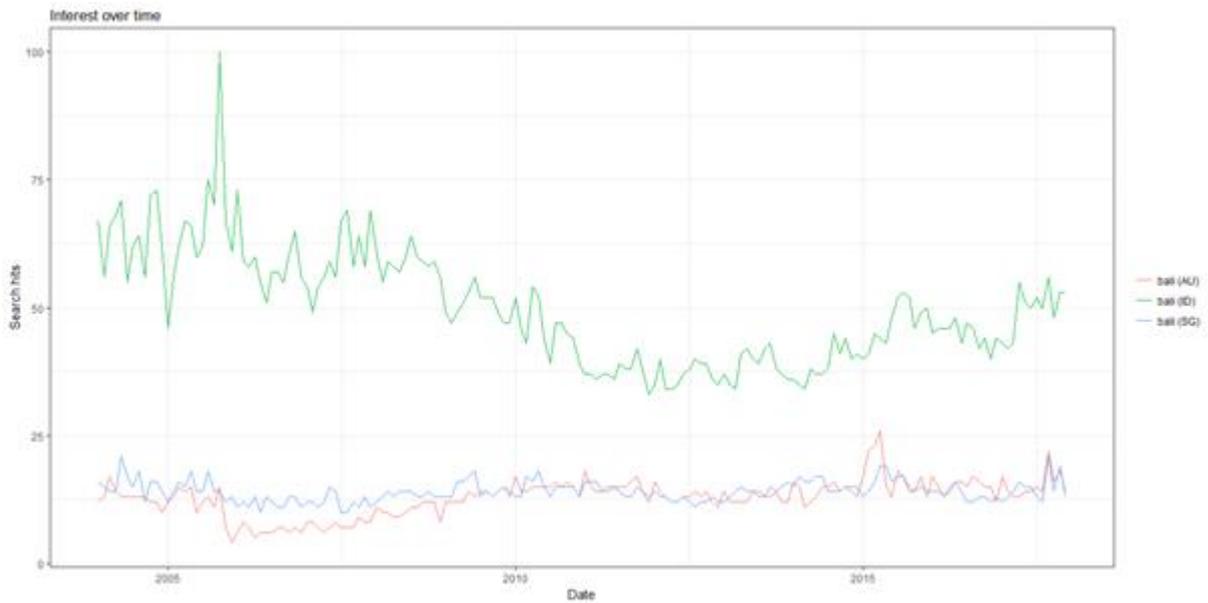

Figure 31. Plot of "Bali" in Australia (AU), Indonesia (ID), and Singapore (SG) for All Categories in Web Search, 2004-2017

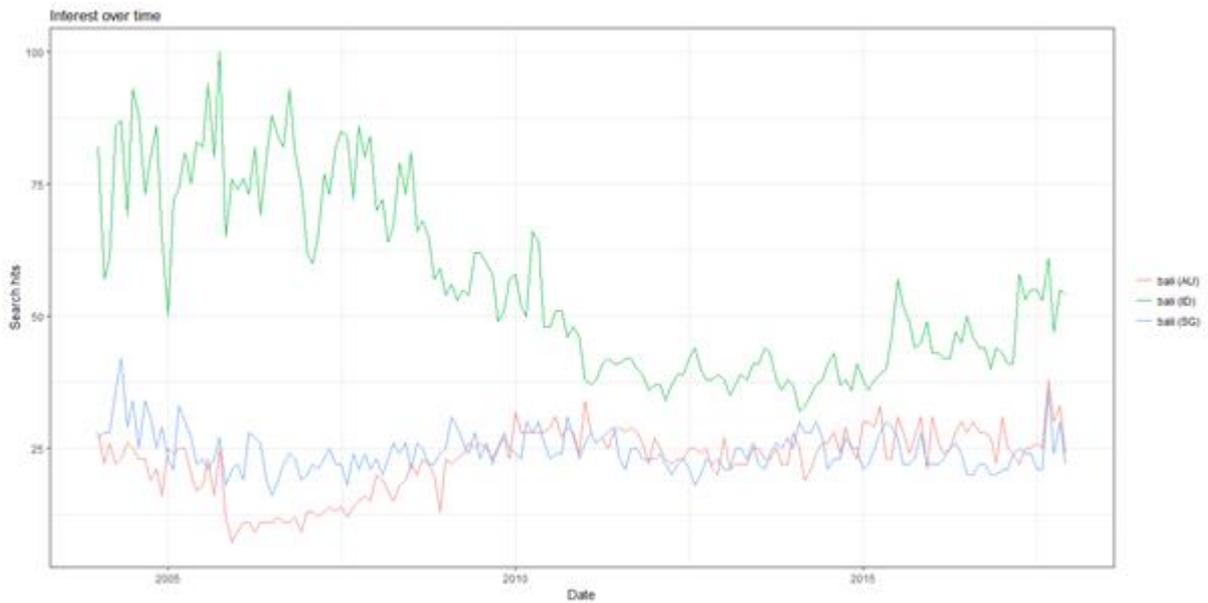

Figure 32. Plot of "Bali" in Australia (AU), Indonesia (ID), and Singapore (SG) for Travel Category in Web Search, 2004-2017

Appendix 3

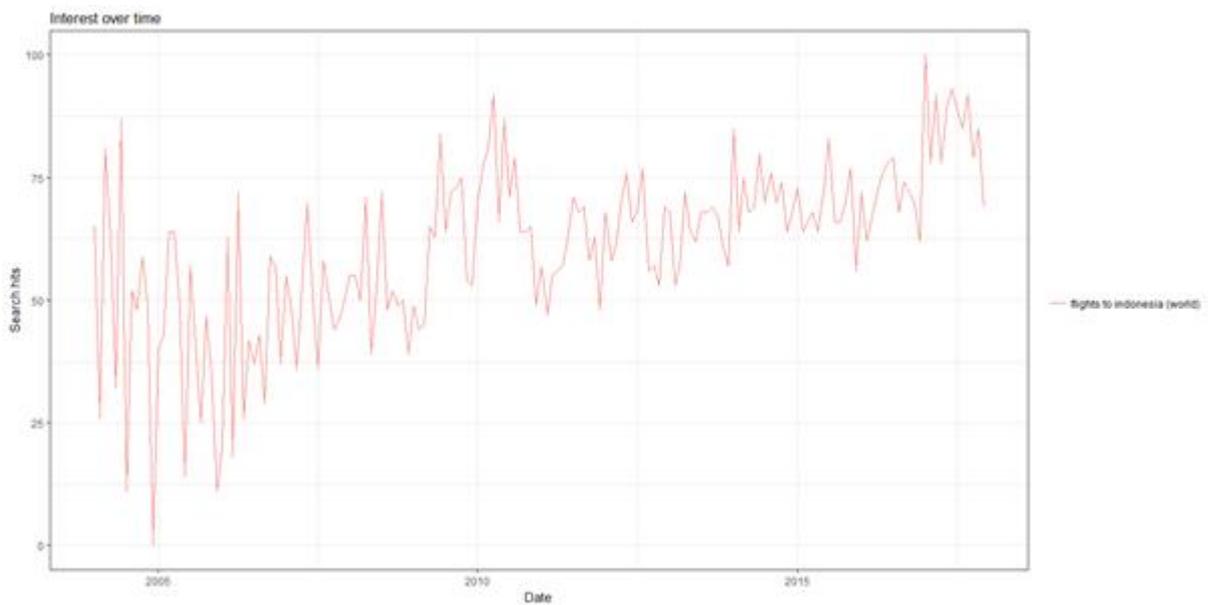

Figure 33. Plot of "flights to Indonesia" for All Categories in Web Search, 2004-2017

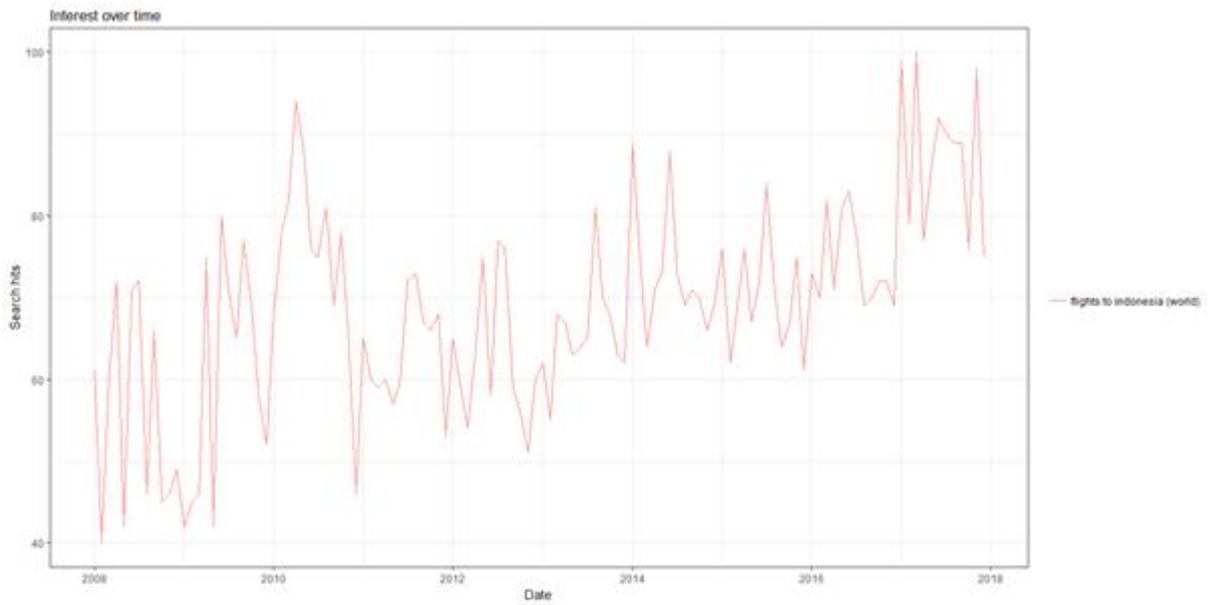

Figure 34. Plot of "flights to Indonesia" for All Categories in Web Search, 2008-2017

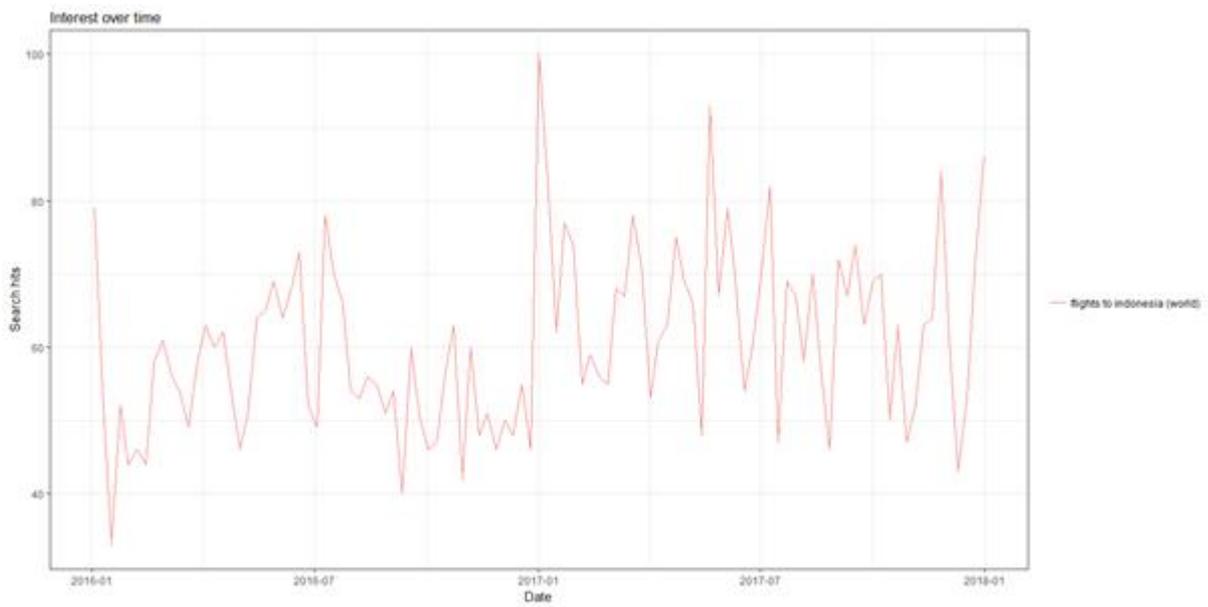

Figure 35. Plot of "flights to Indonesia" for All Categories in Web Search, 2016-2017

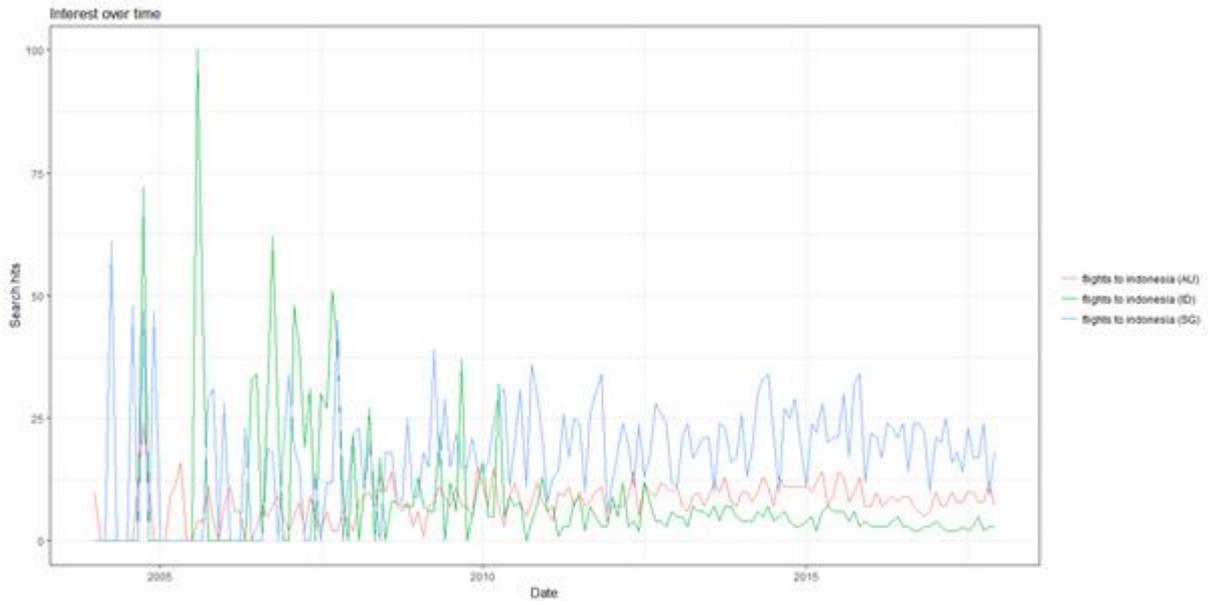

Figure 36. Plot of "flights to Indonesia" in Australia (AU), Indonesia (ID), and Singapore (SG) for All Categories in Web Search, 2004-2017

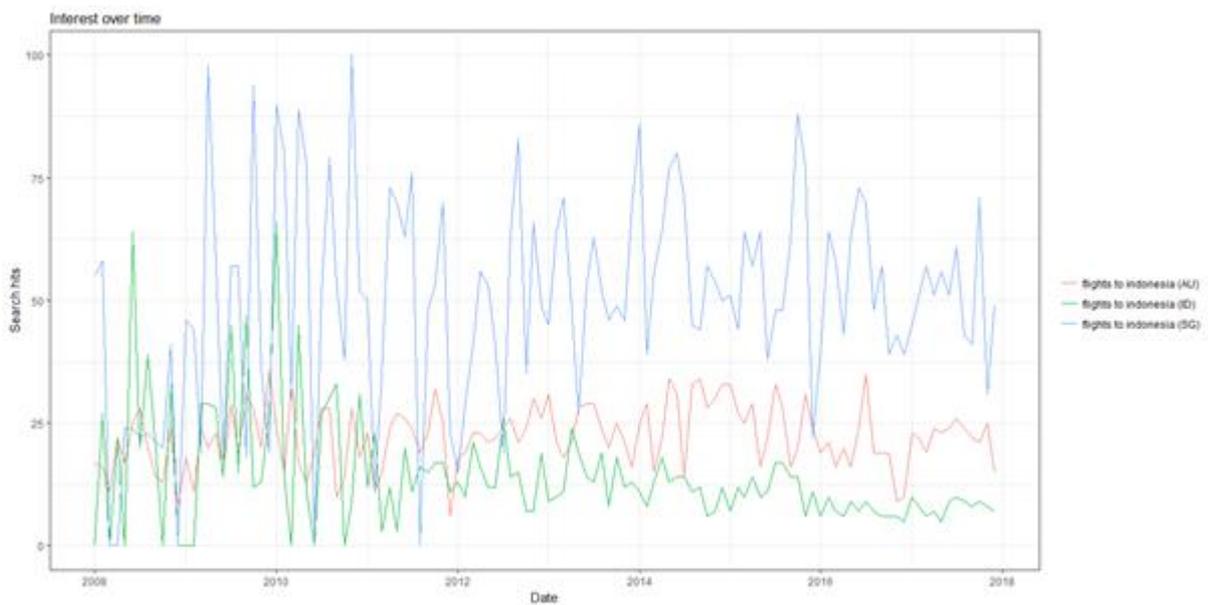

Figure 37. Plot of "flights to Indonesia" in Australia (AU), Indonesia (ID), and Singapore (SG) for All Categories in Web Search, 2008-2017

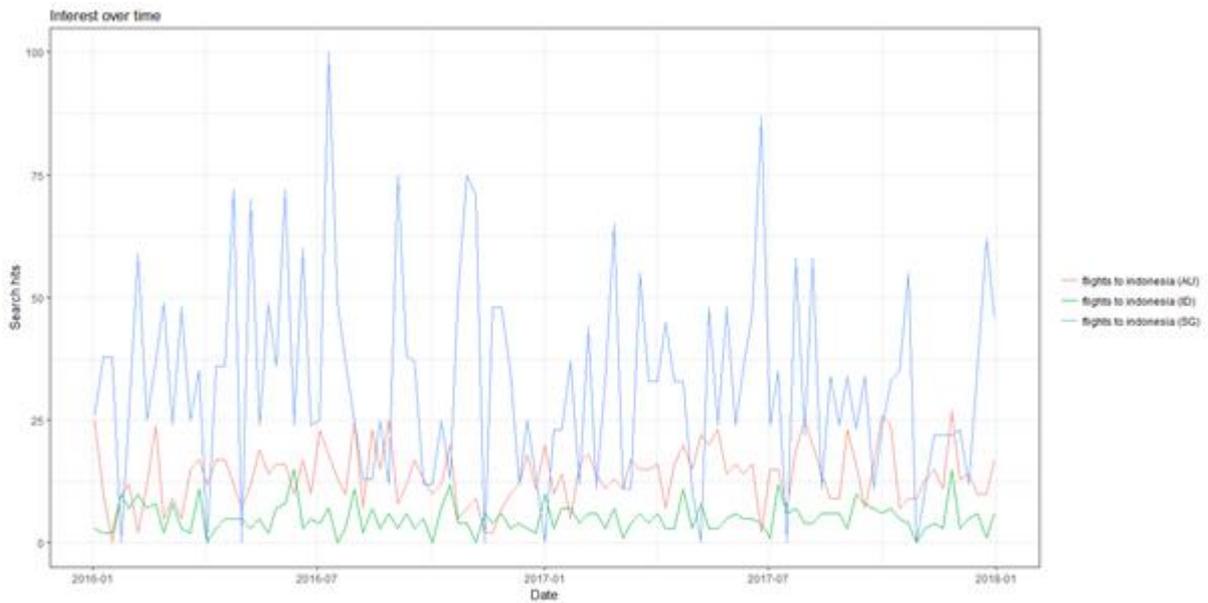

Figure 38. Plot of "flights to Indonesia" in Australia (AU), Indonesia (ID), and Singapore (SG) for All Categories in Web Search, 2016-2017

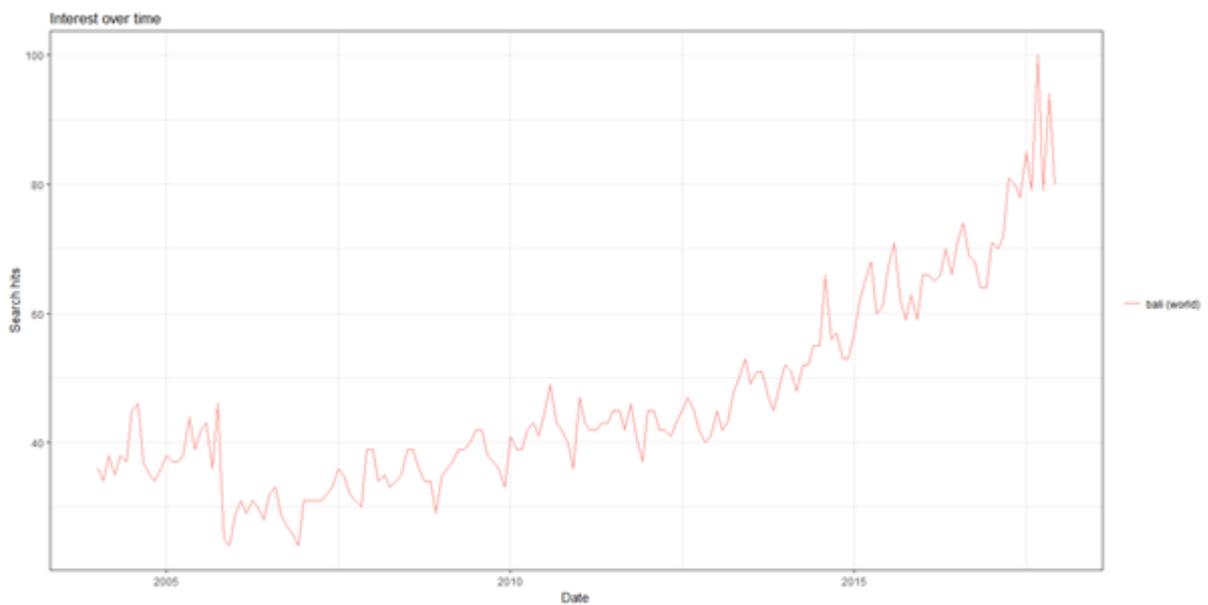

Figure 39. Plot of "Bali" for All Categories in Web Search, 2004-2017

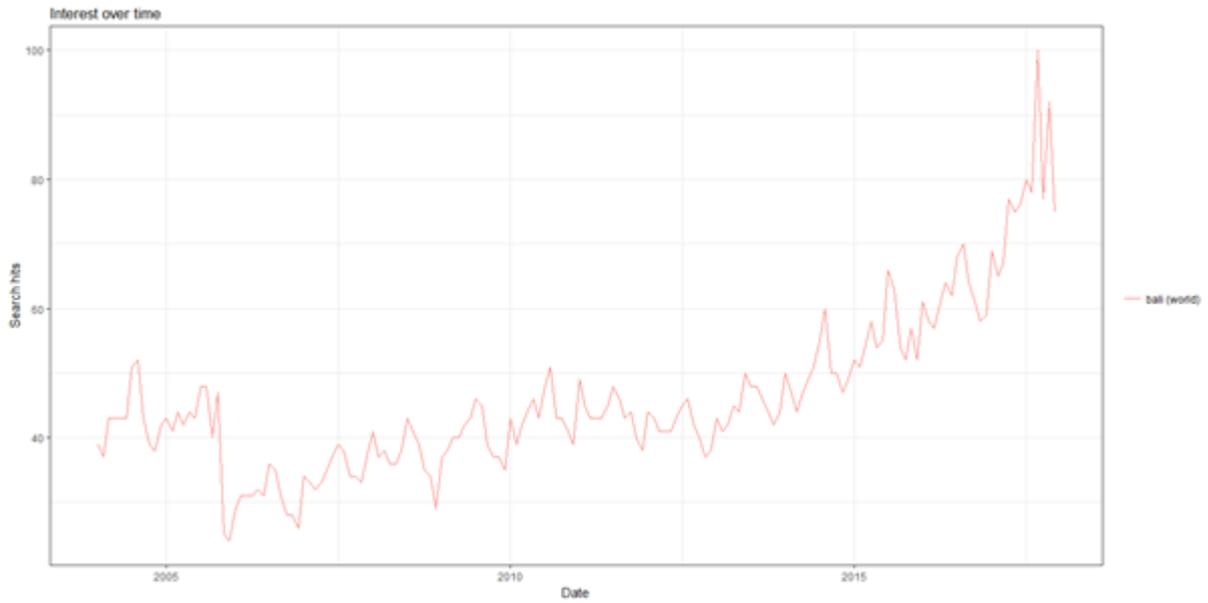

Figure 40. Plot of "Bali" for Travel Category in Web Search, 2004-2017

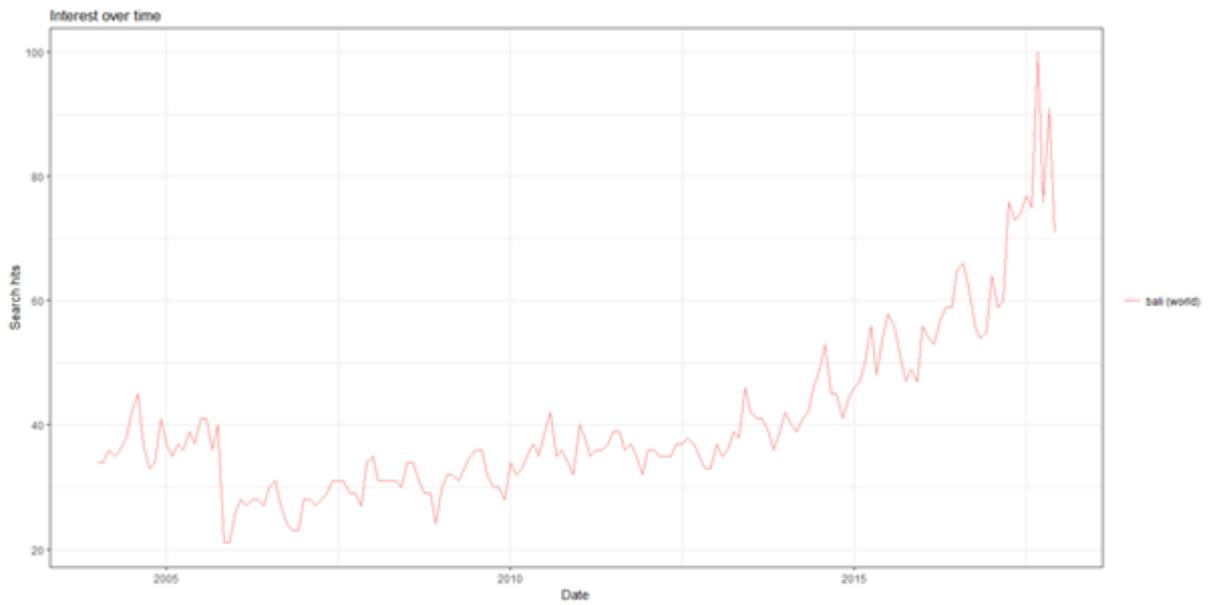

Figure 41. Plot of "Bali" for Tourist Destination Category in Web Search, 2004-2017

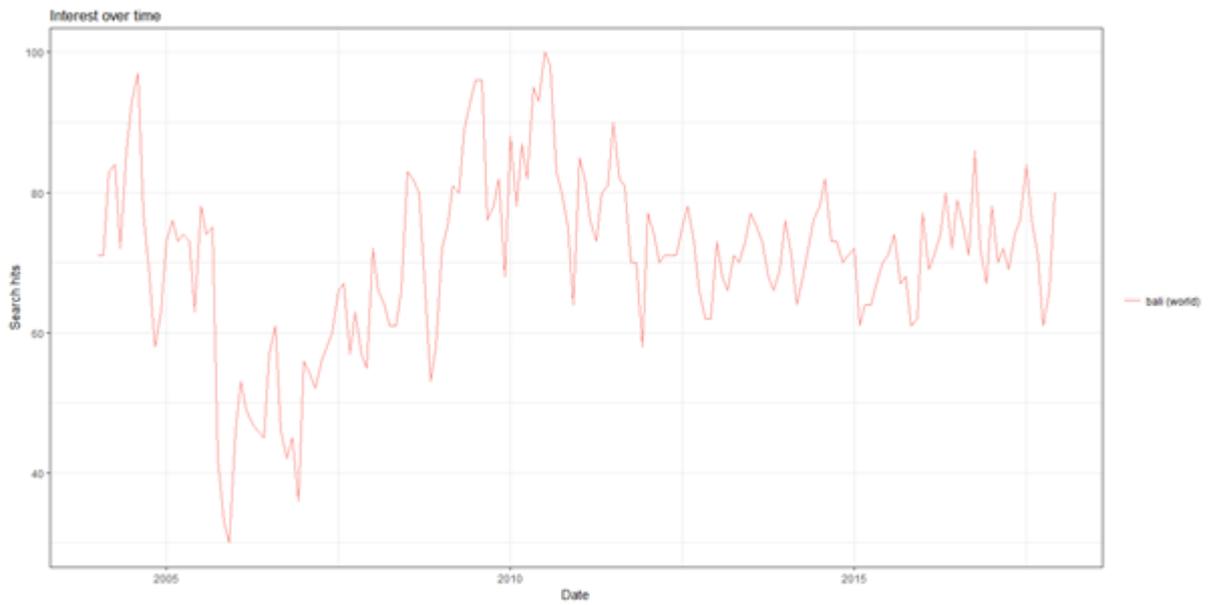

Figure 42. Plot of "Bali" for Hotels and Accommodations Category in Web Search, 2004-2017

Appendix 4

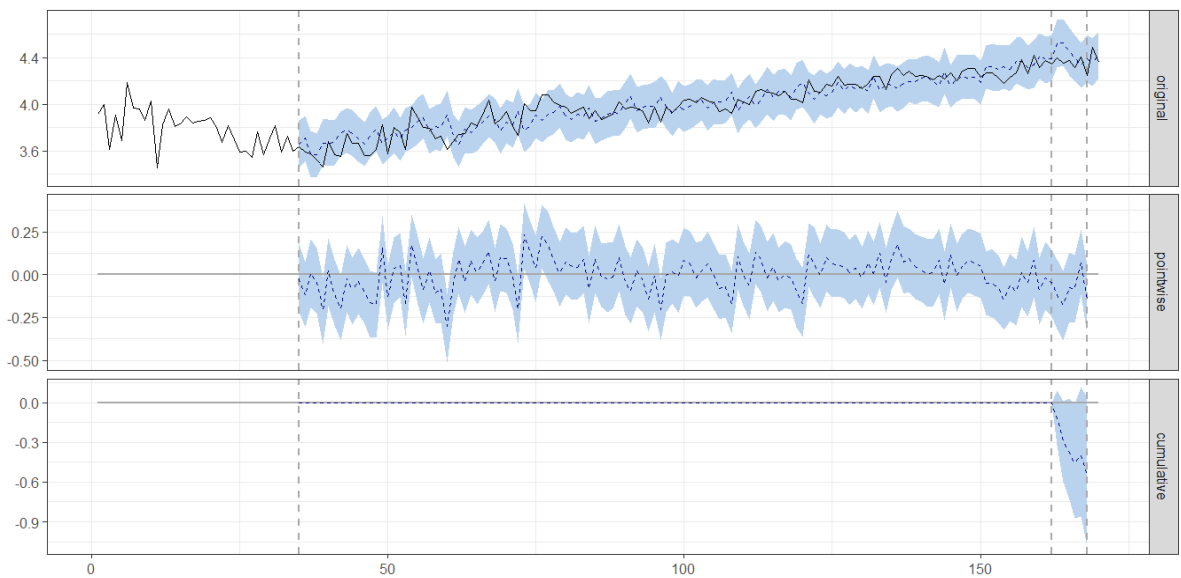

Figure 43. Inferring Causal Impact of Volcanic Eruption in Bali which Happened in 2017

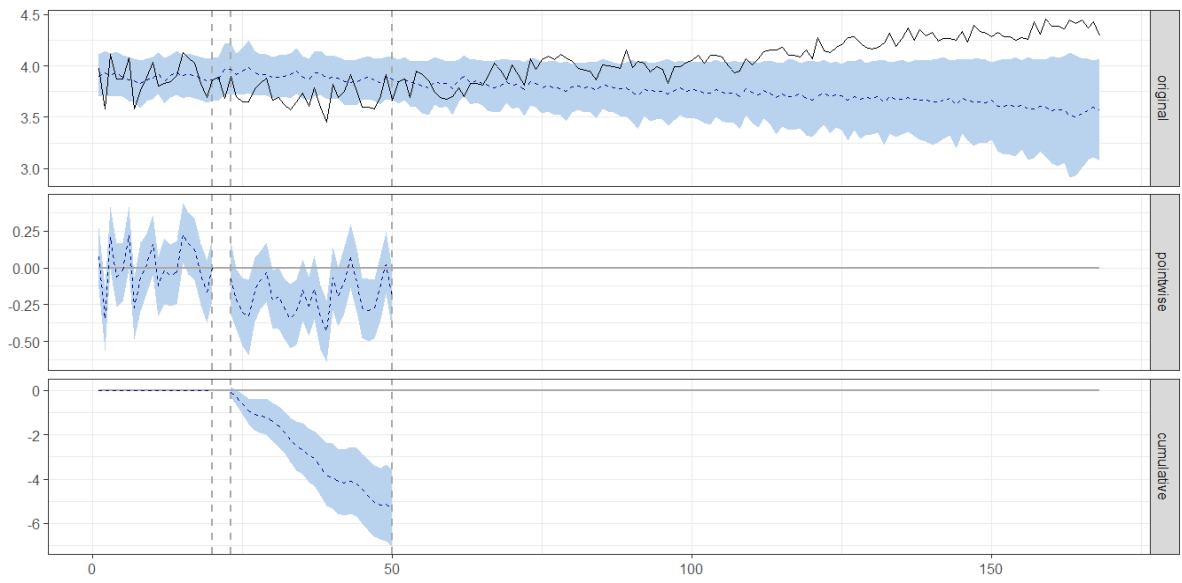

Figure 44. Inferring Causal Impact of Bali Bombing which Happened in 2005